\documentclass[aps,pra,reprint,superscriptaddress,amsmath,amssymb]{revtex4-2}

\usepackage{graphicx}
\usepackage{longtable}
\usepackage{xcolor}
\usepackage[bookmarks,bookmarksnumbered,colorlinks,allcolors=blue]{hyperref}
\usepackage{bookmark}
\usepackage{ifthen}
\usepackage{ulem}

\definecolor{dkgreen}{rgb}{0,0.6,0}
\definecolor{gray}{rgb}{0.5,0.5,0.5}
\definecolor{mauve}{rgb}{0.58,0,0.82}

\newcommand{\ket}[1]{|{#1}\rangle}

\def\be#1\ee{\begin{equation}#1\end{equation}}
\def\ba#1\ea{\begin{align}#1\end{align}}
\def\bg#1\eg{\begin{gather}#1\end{gather}}
\def\t{\text}

\begin{document}

\title{Fluxonium: an alternative qubit platform for high-fidelity operations}

\author{Feng Bao}
\affiliation{Alibaba Quantum Laboratory, Alibaba Group, Hangzhou, Zhejiang 311121, P.R.China}
\author{Hao Deng}
\affiliation{Alibaba Quantum Laboratory, Alibaba Group, Hangzhou, Zhejiang 311121, P.R.China}
\author{Dawei Ding}
\affiliation{Alibaba Quantum Laboratory, Alibaba Group USA, Bellevue, WA 98004, USA}
\author{Ran Gao}
\affiliation{Alibaba Quantum Laboratory, Alibaba Group, Hangzhou, Zhejiang 311121, P.R.China}
\author{Xun Gao}
\affiliation{Alibaba Quantum Laboratory, Alibaba Group USA, Bellevue, WA 98004, USA}
\author{Cupjin Huang}
\affiliation{Alibaba Quantum Laboratory, Alibaba Group USA, Bellevue, WA 98004, USA}
\author{Xun Jiang}
\affiliation{Alibaba Quantum Laboratory, Alibaba Group, Hangzhou, Zhejiang 311121, P.R.China}
\author{Hsiang-Sheng Ku}
\affiliation{Alibaba Quantum Laboratory, Alibaba Group, Hangzhou, Zhejiang 311121, P.R.China}
\author{Zhisheng Li}
\affiliation{Alibaba Quantum Laboratory, Alibaba Group, Hangzhou, Zhejiang 311121, P.R.China}
\author{Xizheng Ma}
\affiliation{Alibaba Quantum Laboratory, Alibaba Group, Hangzhou, Zhejiang 311121, P.R.China}
\author{Xiaotong Ni}
\affiliation{Alibaba Quantum Laboratory, Alibaba Group, Hangzhou, Zhejiang 311121, P.R.China}
\author{Jin Qin}
\affiliation{Alibaba Quantum Laboratory, Alibaba Group, Hangzhou, Zhejiang 311121, P.R.China}
\author{Zhijun Song}
\affiliation{Alibaba Quantum Laboratory, Alibaba Group, Hangzhou, Zhejiang 311121, P.R.China}
\author{Hantao Sun}
\affiliation{Alibaba Quantum Laboratory, Alibaba Group, Hangzhou, Zhejiang 311121, P.R.China}
\author{Chengchun Tang}
\affiliation{Alibaba Quantum Laboratory, Alibaba Group, Hangzhou, Zhejiang 311121, P.R.China}
\author{Tenghui Wang}
\affiliation{Alibaba Quantum Laboratory, Alibaba Group, Hangzhou, Zhejiang 311121, P.R.China}
\author{Feng Wu}
\affiliation{Alibaba Quantum Laboratory, Alibaba Group, Hangzhou, Zhejiang 311121, P.R.China}
\author{Tian Xia}
\affiliation{Alibaba Quantum Laboratory, Alibaba Group, Hangzhou, Zhejiang 311121, P.R.China}
\author{Wenlong Yu}
\affiliation{Alibaba Quantum Laboratory, Alibaba Group, Hangzhou, Zhejiang 311121, P.R.China}
\author{Fang Zhang}
\affiliation{Alibaba Quantum Laboratory, Alibaba Group USA, Bellevue, WA 98004, USA}
\author{Gengyan Zhang}
\affiliation{Alibaba Quantum Laboratory, Alibaba Group, Hangzhou, Zhejiang 311121, P.R.China}
\author{Xiaohang Zhang}
\affiliation{Alibaba Quantum Laboratory, Alibaba Group, Hangzhou, Zhejiang 311121, P.R.China}
\author{Jingwei Zhou}
\affiliation{Alibaba Quantum Laboratory, Alibaba Group, Hangzhou, Zhejiang 311121, P.R.China}
\author{Xing Zhu}
\affiliation{Alibaba Quantum Laboratory, Alibaba Group, Hangzhou, Zhejiang 311121, P.R.China}
\author{Yaoyun Shi}
\email{y.shi@alibaba-inc.com}
\affiliation{Alibaba Quantum Laboratory, Alibaba Group USA, Bellevue, WA 98004, USA}
\author{Jianxin Chen}
\affiliation{Alibaba Quantum Laboratory, Alibaba Group USA, Bellevue, WA 98004, USA}
\author{Hui-Hai Zhao}
\affiliation{Alibaba Quantum Laboratory, Alibaba Group, Beijing 100102, P.R.China}
\author{Chunqing Deng}
\email{chunqing.cd@alibaba-inc.com}
\affiliation{Alibaba Quantum Laboratory, Alibaba Group, Hangzhou, Zhejiang 311121, P.R.China}

\begin{abstract}
  Superconducting qubits provide a promising path toward building large-scale quantum computers. The simple and robust transmon qubit has been the leading platform, achieving multiple milestones. However, fault-tolerant quantum computing calls for qubit operations at error rates significantly lower than those exhibited in the state of the art. Consequently, alternative superconducting qubits with better error protection have attracted increasing interest. Among them, fluxonium is a particularly promising candidate, featuring large anharmonicity and long coherence times. Here, we engineer a fluxonium-based quantum processor that integrates high qubit-coherence, fast frequency-tunability, and individual-qubit addressability for reset, readout, and gates. With simple and fast gate schemes, we achieve an average single-qubit gate fidelity of $99.97\%$ and a two-qubit gate fidelity of up to $99.72\%$. This performance is comparable to the highest values reported in the literature of superconducting circuits. Thus our work, for the first time within the realm of superconducting qubits, reveals an approach toward fault-tolerant quantum computing that is alternative and competitive to the transmon system.   
\end{abstract}
\maketitle

\bookmarksetup{startatroot}

The performance of a quantum processor, characterized by the fidelity of its operations, is confined by the ratio between the operation time and the decoherence times of the qubits. Early superconducting qubits, such as the Cooper-pair box qubit~\cite{nakamura1999coherent} and flux qubit~\cite{mooij1999josephson}, suffer from extremely fast dephasing rates due to their large susceptibility to charge or flux noises. By shunting the Josephson junction with a large capacitor, the transmon qubit~\cite{koch2007charge} protects against charge-noise-induced dephasing, leading to its exceptional success in the past decade~\cite{arute2019quantum, chen2021exponential, kandala2017hardware}. However, this protection comes at the cost of a reduced anharmonicity. Although advanced qubit control schemes~\cite{motzoi2009simple, negirneac2021high} and frequency-selective processor architectures~\cite{chamberland2020topological, hertzberg2021laser} can be adopted to circumvent the speed limit of gate operations imposed by the weak anharmonicity, the leakage to noncomputational states in such fast operations remains a challenge to quantum error correction~\cite{varbanov2020leakage, mcewen2021removing}. Furthermore, transmon qubits still suffer from relaxation due to material imperfections such as dielectric loss~\cite{barends2013coherent, wang2015surface}, which can be challenging to improve \cite{deleon2021materials}.

In contrast, by shunting the Josephson junction with a large linear inductor, fluxonium~\cite{manucharyan2009fluxonium} protects against both charge- and flux- induced dephasing while retaining a large anharmonicity. Additionally, energy relaxation is largely reduced in fluxonium due to its suppressed coupling to dielectric loss. Indeed, isolated single fluxonium qubits have demonstrated long coherence times ranging from a few hundred microseconds to a millisecond~\cite{nguyen2019high, zhang2021universal, somoroff2021millisecond}. 

Despite these intuitive advantages, combining high coherence with fast operations, including gates, reset and readout, in a single high-fidelity fluxonium processor remains challenging. First, the large shunt inductor, composed by more than a hundred Josephson junctions, is susceptible to additional decoherence sources. The best fluxonium qubits achieves an intrinsic loss tangent of approximately $10^{-6}$~\cite{nguyen2019high, somoroff2021millisecond}, defined as the qubit decay rate divided by its frequency, an order of magnitude worse than that achieved in the state-of-the-art transmon~\cite{place2021new, wang2021transmon}. Therefore, fluxonium qubits need to be operated at a much lower frequency than a typical transmon qubit to reach a comparable coherence time. Second, the suppression of energy relaxation at a low qubit frequency is often accompanied by a slowdown of the two-qubit gate speed, limited by the direct qubit-qubit coupling strength. Consequently, two-qubit gate demonstrations~\cite{ficheux2021fast, xiong2021arbitrary} thus far have been limited to microwave-activated schemes that exploit dynamics outside the computational space and thus are limited by the relatively short-lived higher excited states. 

\begin{figure*}
\includegraphics{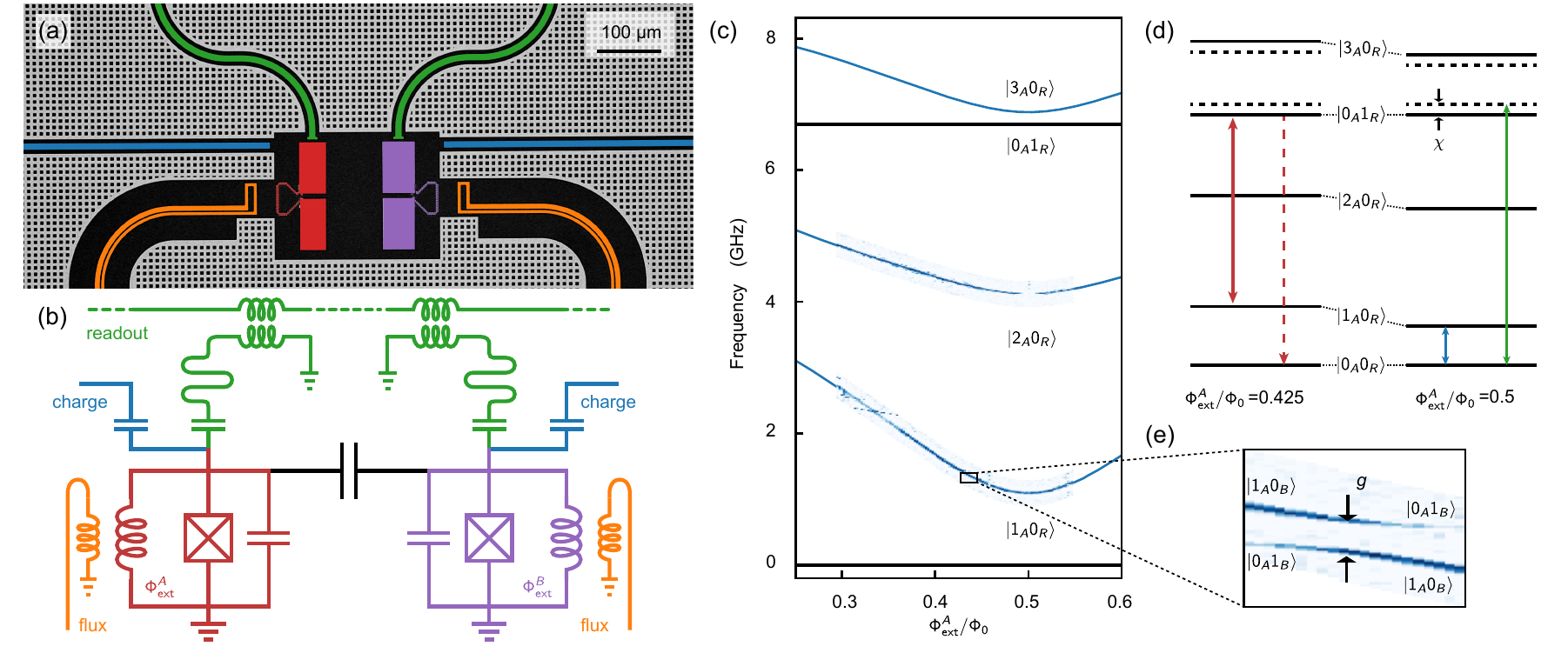}
\caption{\label{fig:layout}Processor architecture. (a) False-colored optical image of the fluxonium processor made of aluminum (colored and white) on a silicon substrate (black). The colored metals correspond to the circuit components in the schematic shown in (b). The two qubits $Q_A$ (red) and $Q_B$ (purple), each consisting of a superconducting loop made from Josephson junctions (see Supplementary Information for a magnified image) and a shunting capacitor, couple to each other through direct capacitive coupling. Individually, qubits are driven through capacitively coupled microwave charge lines (blue), frequency control is achieved through inductively coupled flux lines (orange), and readout is achieved by measuring the resonant frequency shift of the microwave resonators (green). (c) Spectroscopy of $Q_A$ as a function of the external flux through the qubit loop. Lines indicate a fit to the spectrum of the qubit model. (d) Energy diagrams of the qubit-resonator coupled system for $Q_A$ at the $\Phi_\t{ext}^A = 0.5\Phi_0$ flux sweet spot and $\Phi_\t{ext}^A = 0.425\Phi_0$ where the reset is performed. Microwave irradiations used for qubit excitation (blue), reset (red), and readout (green) are shown as arrows. Dashed lines indicate levels without qubit-resonator interaction. (e) Two-qubit spectroscopy with the transition frequency of $Q_A$ sweeping across that of $Q_B$ idled at the flux sweet spot.}
\end{figure*}

We perform meticulous engineering to combine high-coherence fluxonium qubits with a planar integrated circuit that is similar to a scalable transmon processor~\cite{barends2014superconducting}. As shown in Fig.~\ref{fig:layout}(a) and (b), our processor consists of two capacitively-coupled fluxonium qubits $Q_A$ and $Q_B$ in a circuit quantum electrodynamics architecture~\cite{blais2021circuit}. The qubits can be addressed individually using microwave signals to perform independent excitation through charge lines and dispersive-readout through dedicated quarter-wave resonators. Using on-chip flux lines, we independently control the frequency of each qubit by imposing a flux $\Phi_\t{ext}^\alpha$ ($\alpha = A\text{ or }B$) that threads its qubit loop, formed between the phase-slip Josephson junction and the linear inductor consisting of an array of large junctions. Importantly, the large bandwidth of the flux lines allows for rapid tuning of the qubit frequency. This enables operating the qubit at different frequencies for reset, gates, and readout for best performance as well as two-qubit swapping operations that remain entirely within the computational space at a theoretically-maximum speed limited by the qubit-qubit coupling strength.

The measured qubit spectrum versus the flux bias of $Q_A$ and its fit to the fluxonium model is shown in Fig.~\ref{fig:layout}(c) (see Supplementary Information for the full set of extracted device parameters). At $\Phi_{\text{ext}}^A = \Phi_0/2$, the qubit frequency, defined as the transition frequency between the ground and first excited states, is first-order insensitive to flux noise (known as the flux sweet spot) and $\omega_{10} = 2\pi\times 1.09$~GHz. The transition frequency between the first and second excited states is given by $\omega_{21} = 2\pi\times 3.02$~GHz. The qubit anharmonicity, measured by a quantity defined as $(\omega_{21}-\omega_{10})/\omega_{10}$, is 1.771, one order of magnitude larger than that of a typical transmon qubit. We measure the qubit-qubit coupling strength by probing the flux-dependent spectrum of $Q_A$ with $Q_B$ idled at its flux sweet spot, at a qubit frequency of 1.33~GHz (Fig.~\ref{fig:layout}(e)). From the level repulsion due to transverse coupling between the states $\ket{1_A 0_B}$ and $\ket{0_A 1_B}$, where $\ket{k_A l_B}$ denotes excitation level $k(l)$ for $Q_A(Q_B$), a spin-exchange interaction strength $g = 2\pi\times 11.2$~MHz is obtained from the minimum size of the energy splitting.

To exploit the large anharmonicity and the first-order protection against flux noise of fluxonium, we operate the qubits around their $\Phi_{\text{ext}}^A =\Phi_{\text{ext}}^B = \Phi_0/2$ flux sweet spots. In Fig.~\ref{fig:layout}(d), we show the energy diagram of the coupled qubit-resonator system for $Q_A$ to explain the resonator-assisted operations, including the readout and reset. The level labeled $\ket{k_A n_R}$ corresponds to a product state with $k_A$ and $n_R$ excitation in the qubit $Q_A$ and in the resonator, respectively. Although the qubit is far detuned from the readout resonator with a resonant frequency $\omega_R = 2\pi\times 6.696$~GHz, a sizable qubit state dependent resonance shift $\chi = 2\pi\times 0.63$~MHz is obtained and can be attributed to the strong coupling and hybridization of the $\ket{3_A 0_R}$ and $\ket{0_A 1_R}$ states (see Supplementary Information). Thus, qubit readout can be achieved by performing homodyne measurements at the resonator frequency. The readout resonator has a photon spontaneous emission rate $\kappa \approx (70~\t{ns})^{-1}$, which allows for fast readout, and is also used as a damping channel for qubit reset. We note that reset is a necessary operation for fluxonium as the qubit energy $\hbar\omega_{10} \lesssim k_B T$ therefore the first excited state can be thermally populated. We implemented a red-sideband reset similar to Ref.~\cite{manucharyan2009coherent}, in which the population is first transferred from $\ket{1_A 0_R}$ to $\ket{0_A 1_R}$ and then quickly relaxes to the ground state $\ket{0_A 0_R}$ through resonator emission. Simultaneous to the red-sideband drive, we apply a fast flux pulse to offset the qubit away from the $\Phi_\t{ext}^A = \Phi_0/2$ point, lifting the selection rule for this sideband transition. This reset scheme is naturally built in our processor architecture. It is resource efficient and robust in a sense that it requires neither multiple microwave tones nor high-power excitations that could lead to spurious transitions thus is advantageous over the previous schemes~\cite{manucharyan2009coherent, zhang2021universal}. After reset, we identified a ground state population of greater than $95\%$ and a readout contrast of $88\%$. The reset and readout fidelity can be further improved by increasing the side-band transition matrix element with larger flux offset and increasing the readout signal-to-noise ratio with a quantum limited amplifier~\cite{gusenkova2021quantum}.

The coherence times of the qubits at the flux sweet spots are measured to be $T_1=80\, (57)$~$\mu$s and $T_{2,\text{Echo}}=30\,(17)$~$\mu$s for $Q_A$($Q_B$). We also measured the flux dependence of coherence times within a few hundred MHz qubit frequency in the vicinity of the $\Phi_\t{ext}^\alpha = \Phi_0/2$ point where both $T_1$ and $T_2$ nominally have the highest values. The data is consistent with dielectric loss, \textit{i.e.,} relaxation to two-level systems, and dephasing due to low-frequency flux noise (see Supplementary Information).

As described, we have built a quantum computing processor that utilizes the large anharmonic spectrum of fluxonium. The lowest two levels used as the qubit with a small energy gap are reasonably well-protected from decoherence while remaining easily addressable by resonant drive and allowing a strong qubit-qubit coupling strength. Reset and readout operations require strong interactions with the environment and are realized through higher, noncomputational levels. Together with fast qubit-frequency tunability, our scheme combines high qubit coherence with fast gates, reset, and readout. In addition, this processor architecture requires no more on-chip resources than a transmon processor~\cite{barends2014superconducting}, which has been demonstrated as a scalable technology toward dozens to hundreds of qubits. 

\begin{figure}
  \includegraphics{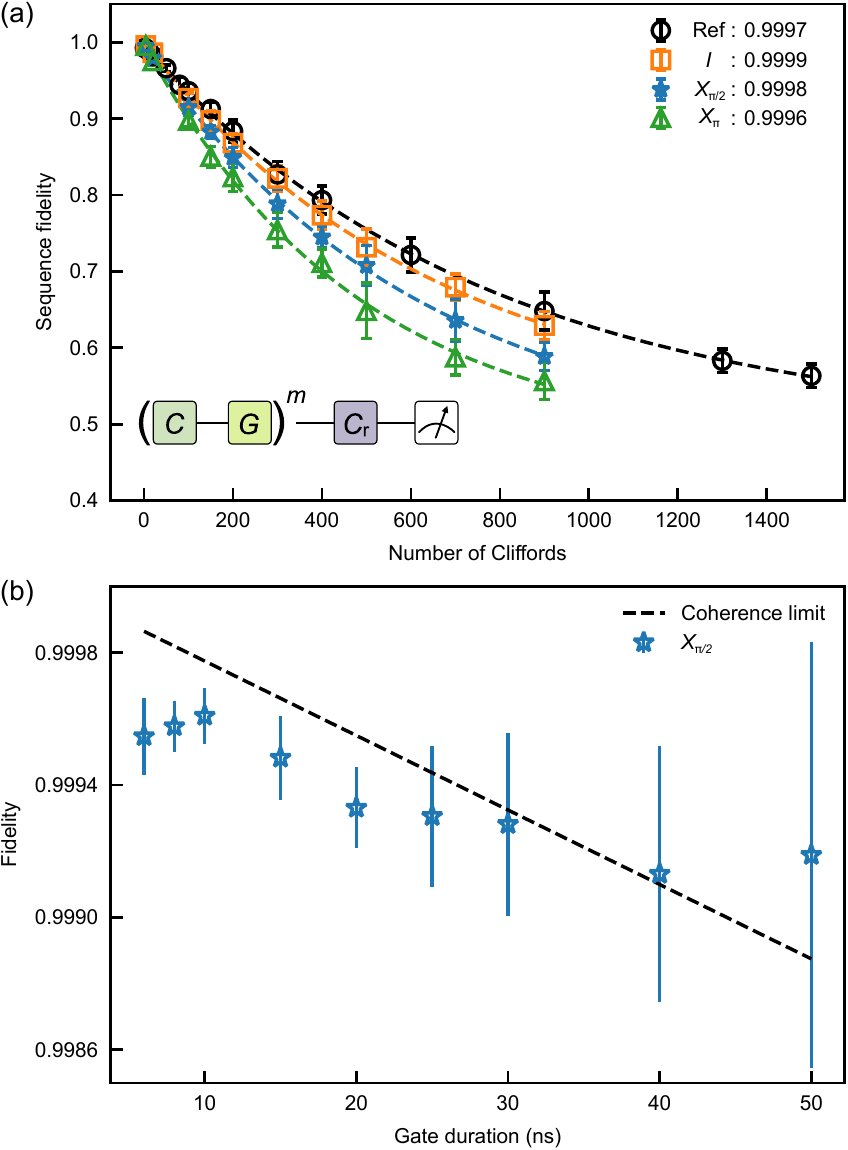}
  \caption{\label{fig:1q_gate}Single-qubit gate and randomized benchmarking (RB). (a) Sequence fidelities of reference and interleaved RB with sequences of $m$ random Clifford gates. For the fidelity of each value of $m$, 20 random sequences are used to obtain the average sequence fidelity and the standard deviation is displayed as error bars. The insert represents the gate sequence of interleaved RB for the characterization of a specific gate $G$. For standard RB, the specific gate $G$ is removed from the sequence.  (b) Fidelity of $X_{\pi/2}$ as a function of gate duration. The black dashed line represents the calculated coherence limit.}
  \end{figure}

Because of the large anharmonicity and high coherence, we can perform fast, high-fidelity single qubit operations with simple control pulses. Specifically, we pulse on the resonant microwave drives with simple cosine envelopes (see Supplementary Information) to rotate the qubit in its Bloch sphere. With a fixed gate duration of 10~ns, the angle of the rotation is determined by the amplitude of the drive, and the rotational axis is set by its phase. We calibrate (see Supplementary Information) a primary set of gate operations in qubit $Q_A$, denoted as $\{I, X_{\pi},Y_{\pi},X_{\pm \pi/2},Y_{\pm \pi/2}\}$, consisting of $0$, $\pi$ and $\pi/2$ rotations around two independent axes $X$ and $Y$. 
The fidelity of these primary gates can be characterized using randomized benchmarking (RB)~\cite{knill2008randomized, magesan2011scalable, magesan2012efficient} (Fig.~\ref{fig:1q_gate}(a)). A standard RB circuit consists of $m$ randomly selected Clifford gates ($C$) followed by a recovery gate ($C_r$), such that the full circuit implements an identity operation in the absence of gate errors. However, infidelities in the gate operations, arising from both spontaneous qubit rotations (decoherence error) and imperfections in the control pulses (control error), cause the sequence fidelity, measured as the overlap between the initial and the final state, to deviate from unity. Before every sequence, we initialize the qubit into its ground state by a reset; thus, the sequence fidelity is given by the ground state population measured at the end of the gate sequence. Because the Clifford gates are constructed from a combination of the primary gates, this measurement extracts an average fidelity of $99.97\%$ across all primary gates (see Supplementary Information). We then further benchmark the fidelity of each specific gate ($G$) by appending it after every Clifford operation. We find that the fidelities of identity ($I$) and the $X_{\pi/2}$ gate is $99.99\%$ and $99.98\%$ respectively, in reasonably good agreement with the limit of decoherence (see Supplementary Information). 
However, the error of the $X_\pi$ gate, extracted from its $99.96\%$ fidelity, is more than double of that of the $X_{\pi/2}$ gate and thus cannot be explained by decoherence alone. 

We next measure the gate fidelity of $X_{\pi/2}$ as a function of the gate duration. A gate with a shorter duration is implemented with a larger amplitude to ensure a fixed $\pi/2$ rotation angle. 
The experiment is performed on $Q_B$ and the results are shown in Fig.~\ref{fig:1q_gate}(b). The gate fidelity generally increases with deceasing gate duration and remains close to the coherence limit (black dashed line). We observe no sign of elevated gate error on short gate durations down to 10~ns without employing advanced pulse-shaping techniques such as derivative removal by adiabatic gate (DRAG)~\cite{motzoi2009simple}. Using numerical simulations, we find population leakage as low as $10^{-8}$ at a 10~ns gate duration, enabled entirely by the large anharmonicity of the fluxonium qubit (see Supplementary Information). This shows that the gate fidelity is possibly affected by other sources of imperfections during strong driving, such as heating or pulse distortions~\cite{gustavsson2013improving}. Going beyond the Clifford gates, we combine these primary operations with virtual-$Z$ gates~\cite{mckay2017efficient} to perform arbitrary single-qubit rotations.

\begin{figure*}
  \includegraphics{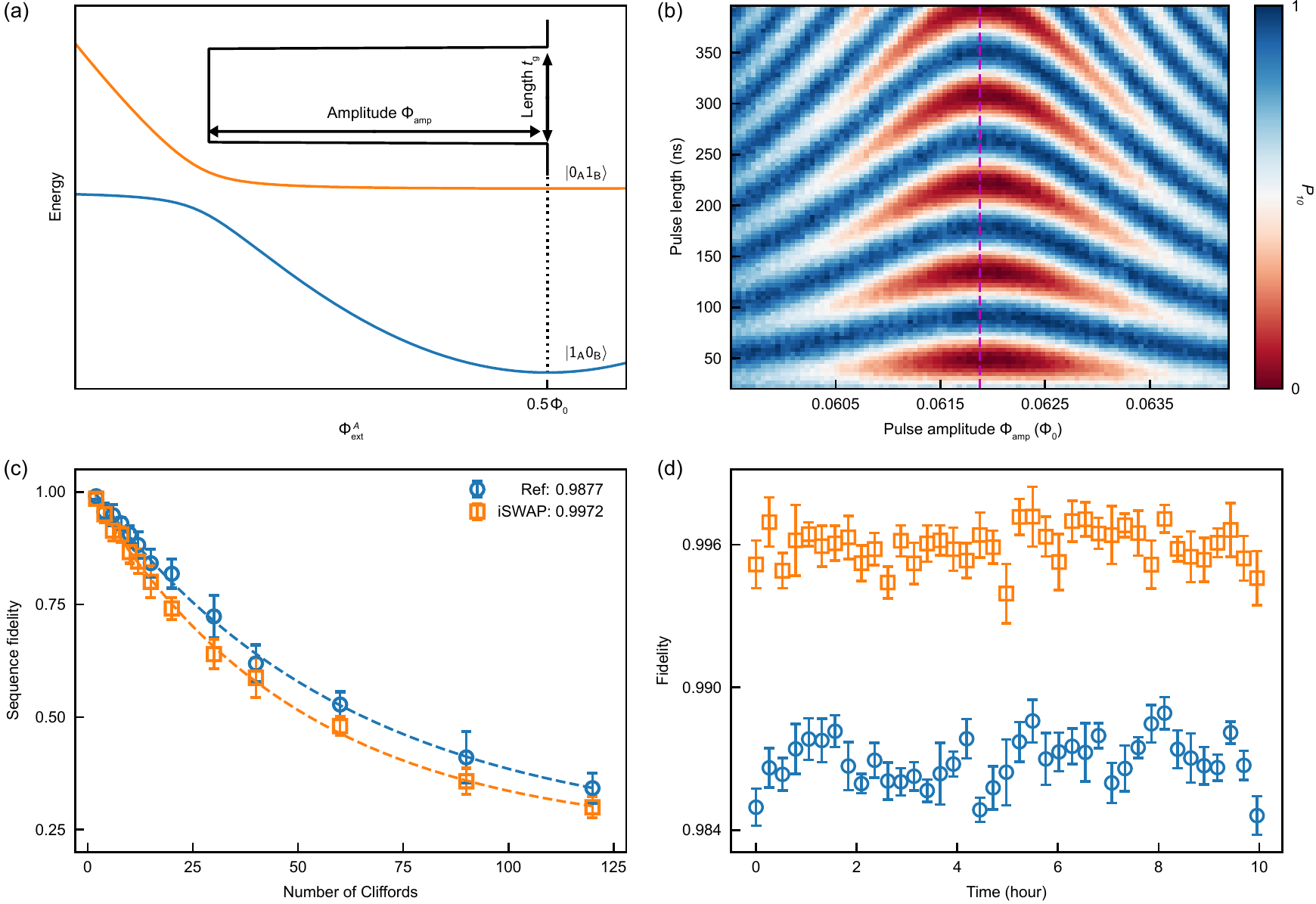}
  \caption{\label{fig:2q_gate}Two-qubit gate scheme and benchmarking results. (a) Schematic of the level structure involving the energy eigenstates $|0_A1_B\rangle$ and $|1_A0_B\rangle$. The insert shows the flux control pulse that activates the iSWAP operation. (b) Coherent oscillation of the population between the state $\ket{0_A1_B}$ and $\ket{1_A0_B}$, shown as $P_{01}$ as a function of the pulse amplitude $\Phi_\t{amp}$ and pulse length $t_{g}$. For each $\Phi_\t{amp}$, a swapping rate can be extracted from a fast Fourier transform of the population oscillation. The magenta dashed line corresponds to the pulse amplitude where the swapping rate is at the minimum and the system undergoes a complete population exchange from one qubit to the other. (c) Sequence fidelity of Clifford RB and interleaved RB of the iSWAP gate. Each fidelity is obtained from averaging over 10 random sequences. (d) Temporal fluctuation of the fidelities for the two-qubit Clifford gates and the iSWAP gate. The error bars are calculated from the uncertainty of fitting parameters.}
  \end{figure*}

To create a universal gate set, we complement our arbitrary single-qubit gates with a two-qubit iSWAP gate. The iSWAP gate is performed at the point of avoided level crossing (Fig.~\ref{fig:2q_gate}(a)) due to the direct qubit-qubit interaction. Specifically, after initializing the qubit pair into $\ket{1_A 0_B}$, we use the broadband flux line to alter the frequency of qubit $Q_A$ and bring it adiabatically into resonance with qubit $Q_B$ for a duration $t_{g}$. Fig.~\ref{fig:2q_gate}(b) shows the resulting spin-exchange-like oscillation between the two qubits at a rate $\sqrt{g^2 + \Delta^2}$ given by the qubit-qubit coupling strength $g \approx 2\pi\times 11.2$~MHz and the qubit frequency difference $\Delta$ set by the flux-pulse amplitude $\Phi_\t{amp}$. An iSWAP gate is realized at $t_g = 50$~ns and $\Delta = 0$ (magenta dashed line), where the initial excitation is swapped entirely into qubit $Q_B$ and the system evolves to a final state $-i\ket{0_A 1_B}$. The fine tuning details of the gate can be found in the Supplementary Information. To characterize the performance of our iSWAP gate, we perform RB as discussed above, but with two-qubit Clifford gates generated by a combination of the iSWAP gate and the single-qubit primary gates, $\{I, X_{\pi},Y_{\pi},X_{\pm \pi/2},Y_{\pm \pi/2}\}$. As shown in Fig.~\ref{fig:2q_gate}(c), we find an average fidelity of $98.77\%$ for two-qubit Clifford gates and a fidelity up to $99.72\%$ for the iSWAP gate. In Fig.~\ref{fig:2q_gate}(d), we show the temporal fluctuation of the gate fidelity. In a 10-hour duration, we find an average iSWAP gate fidelity of $99.60\%$, demonstrating the very good stability of this flux tunable gate. Finally, we analyze the limit of this gate scheme through numerical simulations (see Supplementary Information). We find that 0.05\% infidelity can be attributed to control errors due to pulse imprecision and stray qubit-qubit interactions. Importantly, because of the large anharmonicity of the qubit and that the gate scheme is confined in the computational space, the leakage error is negligible. Instead, our iSWAP gate infidelity is dominated by qubit decoherence, where qubit relaxation and dephasing each contribute $0.06\%$ and $0.26\%$, respectively, to gate errors. 

To summarize, we realize a fluxonium processor with single- and two-qubit gate fidelities comparable to the state of the art in the widely adopted transmon qubits~\cite{sung2021realization, stehlik2021tunable}. This performance is achieved through simple, fast, and low-leakage gate schemes with fast frequency tunability and individual controls. Importantly, this high-fidelity gate set is demonstrated together with built-in robust reset and readout in a single device, thus fulfilling all the criteria of the physical implementation of quantum computing~\cite{divincenzo2000the}.

The theoretical characteristics of fluxonium and our infidelity analyses lead naturally to an avenue discussed below for further developments. Although the fluxonium coherence times in this work have not exceeded those best-reported values, the extracted dielectric loss in our devices is approximately $10^{-6}$ (see Supplementary Information), comparable to the most coherent fluxonium demonstrated thus far~\cite{nguyen2019high, somoroff2021millisecond}. We thus anticipate that a significant improvement in qubit coherence can be achieved by lowering the frequency, without the need of further reducing dielectric loss. This coherence enhancement is a result of the suppressed transition matrix element of the charge operator of fluxonium, which inevitably slows down the gates based on capacitive couplings. Inductive couplings, however, do not suffer from the above limitation. We expect that our gate schemes can be applied to strongly inductively coupled and high-coherence fluxonium qubits, leading to even higher fidelity gates as well as other operations. To scale up, tunable couplers~\cite{arute2019quantum, wu2021strong} are a crucial component to maintain high-fidelity for parallel operations in a large superconducting circuit. Such schemes are known for inductively coupled flux qubits~\cite{harris2007sign, niskanen2007quantum}, which have essentially the same form of Hamiltonian as fluxonium. Therefore, our work not only suggests a viable alternative path for fault-tolerance, but one that may eventually outperform transmon, the current mainstream qubit of choice.


\section*{Acknowledgements}
We thank all full-time associates at Alibaba Quantum Laboratory for experimental  support, Xin Wan and Shi-Biao Zheng for insightful discussions. Y.~S. is indebted to Jeff~Zhang for his patience and support, which is indispensable for the team's risk-taking spirit.

\bibliographystyle{apsrev4-2}
\bibliography{ref}

\end{document}


\title{Supplementary Information for \\``Fluxonium: an alternative qubit platform for high-fidelity operations''}

\author{Feng Bao}
\affiliation{Alibaba Quantum Laboratory, Alibaba Group, Hangzhou, Zhejiang 311121, P.R.China}
\author{Hao Deng}
\affiliation{Alibaba Quantum Laboratory, Alibaba Group, Hangzhou, Zhejiang 311121, P.R.China}
\author{Dawei Ding}
\affiliation{Alibaba Quantum Laboratory, Alibaba Group USA, Bellevue, WA 98004, USA}
\author{Ran Gao}
\affiliation{Alibaba Quantum Laboratory, Alibaba Group, Hangzhou, Zhejiang 311121, P.R.China}
\author{Xun Gao}
\affiliation{Alibaba Quantum Laboratory, Alibaba Group USA, Bellevue, WA 98004, USA}
\author{Cupjin Huang}
\affiliation{Alibaba Quantum Laboratory, Alibaba Group USA, Bellevue, WA 98004, USA}
\author{Xun Jiang}
\affiliation{Alibaba Quantum Laboratory, Alibaba Group, Hangzhou, Zhejiang 311121, P.R.China}
\author{Hsiang-Sheng Ku}
\affiliation{Alibaba Quantum Laboratory, Alibaba Group, Hangzhou, Zhejiang 311121, P.R.China}
\author{Zhisheng Li}
\affiliation{Alibaba Quantum Laboratory, Alibaba Group, Hangzhou, Zhejiang 311121, P.R.China}
\author{Xizheng Ma}
\affiliation{Alibaba Quantum Laboratory, Alibaba Group, Hangzhou, Zhejiang 311121, P.R.China}
\author{Xiaotong Ni}
\affiliation{Alibaba Quantum Laboratory, Alibaba Group, Hangzhou, Zhejiang 311121, P.R.China}
\author{Jin Qin}
\affiliation{Alibaba Quantum Laboratory, Alibaba Group, Hangzhou, Zhejiang 311121, P.R.China}
\author{Zhijun Song}
\affiliation{Alibaba Quantum Laboratory, Alibaba Group, Hangzhou, Zhejiang 311121, P.R.China}
\author{Hantao Sun}
\affiliation{Alibaba Quantum Laboratory, Alibaba Group, Hangzhou, Zhejiang 311121, P.R.China}
\author{Chengchun Tang}
\affiliation{Alibaba Quantum Laboratory, Alibaba Group, Hangzhou, Zhejiang 311121, P.R.China}
\author{Tenghui Wang}
\affiliation{Alibaba Quantum Laboratory, Alibaba Group, Hangzhou, Zhejiang 311121, P.R.China}
\author{Feng Wu}
\affiliation{Alibaba Quantum Laboratory, Alibaba Group, Hangzhou, Zhejiang 311121, P.R.China}
\author{Tian Xia}
\affiliation{Alibaba Quantum Laboratory, Alibaba Group, Hangzhou, Zhejiang 311121, P.R.China}
\author{Wenlong Yu}
\affiliation{Alibaba Quantum Laboratory, Alibaba Group, Hangzhou, Zhejiang 311121, P.R.China}
\author{Fang Zhang}
\affiliation{Alibaba Quantum Laboratory, Alibaba Group USA, Bellevue, WA 98004, USA}
\author{Gengyan Zhang}
\affiliation{Alibaba Quantum Laboratory, Alibaba Group, Hangzhou, Zhejiang 311121, P.R.China}
\author{Xiaohang Zhang}
\affiliation{Alibaba Quantum Laboratory, Alibaba Group, Hangzhou, Zhejiang 311121, P.R.China}
\author{Jingwei Zhou}
\affiliation{Alibaba Quantum Laboratory, Alibaba Group, Hangzhou, Zhejiang 311121, P.R.China}
\author{Xing Zhu}
\affiliation{Alibaba Quantum Laboratory, Alibaba Group, Hangzhou, Zhejiang 311121, P.R.China}
\author{Yaoyun Shi}
\email{y.shi@alibaba-inc.com}
\affiliation{Alibaba Quantum Laboratory, Alibaba Group USA, Bellevue, WA 98004, USA}
\author{Jianxin Chen}
\affiliation{Alibaba Quantum Laboratory, Alibaba Group USA, Bellevue, WA 98004, USA}
\author{Hui-Hai Zhao}
\affiliation{Alibaba Quantum Laboratory, Alibaba Group, Beijing 100102, P.R.China}
\author{Chunqing Deng}
\email{chunqing.cd@alibaba-inc.com}
\affiliation{Alibaba Quantum Laboratory, Alibaba Group, Hangzhou, Zhejiang 311121, P.R.China}

\maketitle

\bookmarksetup{startatroot}
\section{Fabrication}

\begin{figure}[b]
  \includegraphics{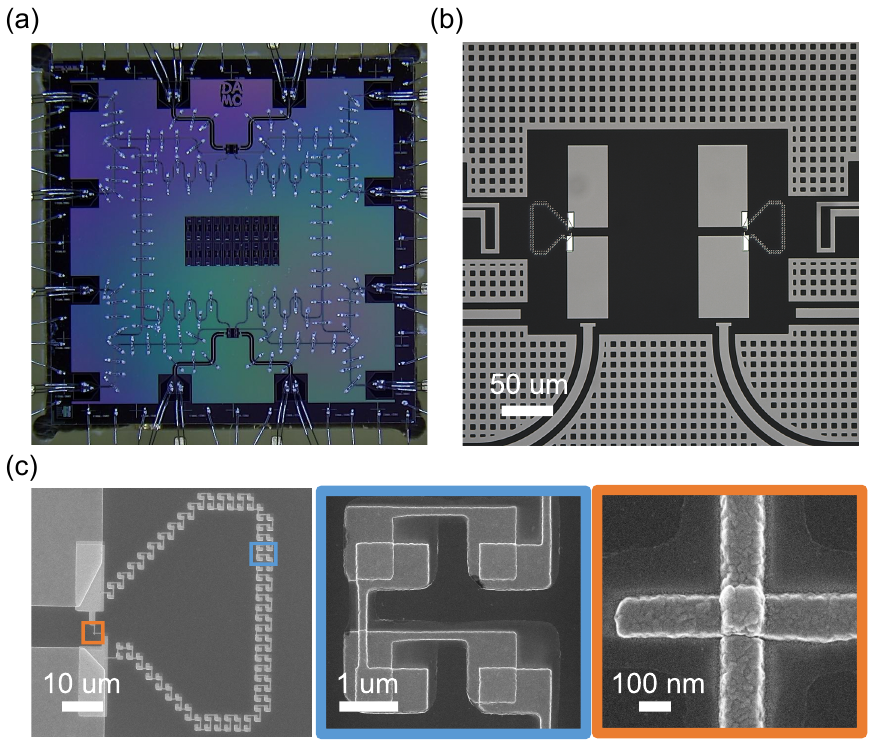}
  \caption{\label{fig:Supp_fig1}(a) The optical image of the wire-bonded fluoxnium processor. (b) The optical image of the region containing two capacitively-coupled qubits. (c) The SEM image of the fluxonium qubit. The zoomed-in images of the junction array and the phase-slip junction are provided.}
\end{figure}

The processor is fabricated on a high-resistivity float-zone silicon substrate manufactured by Topsil GlobalWafers. The substrate is first etched in a 0.5\% hydrofluoric acid bath for 2~\si{\minute} followed by a 2~\si{\minute} de-ionized water rinse at room temperature to remove the native oxides and loaded in an ultrahigh-vacuum cluster of deposition system(AdNanoTek JEB-D) within 5~\si{\minute}. A 150~\si{\nano\metre} thick aluminum film is e-beam evaporated onto the substrate at 1 \si{\nano\metre/\second} deposition rate. The aluminum film is spin coated with a 1~\si{\micro\metre} thick S1813 photoresist (Kayaku Advanced Materials) and patterned by a direct laser writing system (Heidelberg Instruments DWL2000) to form all circuits other than Josephson junctions. The development is performed in a bath of MIF-319 (Kayaku Advanced Materials) for 1~\si{\minute} followed by 3~\si{\minute} de-ionized water rinse at room temperature. A bath of Aluminum Etchant Type A (Transene Company Inc.) at 55~\si{\degreeCelsius} is used to etch the aluminum film for 15~\si{\second} followed by a 3 \si{\minute} de-ionized water rinse. The photoresist is stripped in two baths of acetone, and two baths of isopropyl alcohol (IPA) with 5~\si{\minute} ultrasonication each.

The processed wafer is then spin coated with a bilayer resist consists of 1~\si{\micro\metre} PMMA (poly methyl methacrylate) 950K /200~\si{\nano\metre} MMA (8.5) (methyl methacrylate) and patterned via an e-beam lithography system (JEOL JBX8100FS). The development is performed in a bath of MIBK/IPA 1:3 (Kayaku Advanced Materials) for 1~\si{\minute} with gentle agitation  followed by 1~\si{\minute} IPA rinse. 1~\si{\minute} ion mill is performed to remove the resist residue and aluminum oxide in the structural openings to form a galvanic contact between Josephson junctions and capacitor pads in an ultrahigh-vacuum cluster of deposition system (AdNanoTek JEB-D). The Josephson junctions are then fabricated in a conventional Manhattan-style approach~\cite{Kreikebaum2020}. The 20~\si{\nano\metre} base electrode layer (Al) is e-beam evaporated at 1~\si{\nano\metre/\second} followed by a 40~\si{\min} oxidation at 800~\si{\pascal} to form the barrier of Josephson junctions. The 30~\si{\nano\metre} counter electrode layer (Al) is e-beam evaporated at 1~\si{\nano\metre/\second} followed by a 20~\si{\min} termination oxidation at 2666~\si{\pascal}. The lift-off process is performed in one bath of N-Methyl-2-pyrrolidone at 80~\si{\degreeCelsius} for 2~\si{\hour} followed by one bath of acetone and one bath of IPA with 3~\si{\minute} ultrasonication each. The wafer is diced in a dicing saw system (DISCO DAD3221) with 2~\si{\micro\metre} PMMA 950K protection. The photoresist is stripped in two baths of acetone, two baths of IPA with 5~\si{\minute} ultrasonication each. The processor is wire-bonded in a gold-plated aluminum sample-box for cryogenic measurement. The Optical and SEM images of the devices are shown in Fig.~\ref{fig:Supp_fig1}.

\section{Experimental Setup}
\begin{figure}[htbp]
  \includegraphics{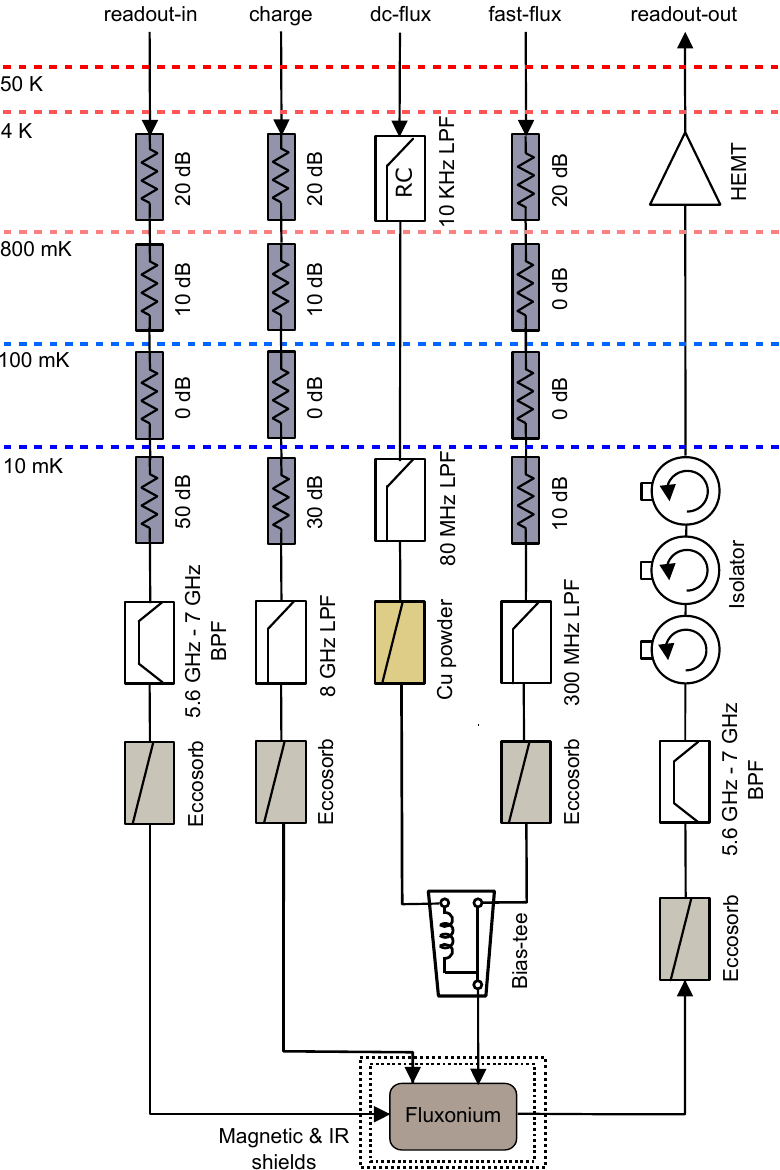}
  \caption{\label{fig:setup} Wiring diagram inside the dilution refrigerator. We only draw the control lines of one qubit and both qubits have identical wiring configurations.}
\end{figure}

As shown in Fig.~\ref{fig:setup}, the processor is cooled down to a base temperature of about $10$~mK in a BlueFors LD-400 dilution refrigerator. The devices are magnetically shielded with a superconducting container surrounded by mu-metals. Control signals for the readout-in, charge and fast-flux lines are generated by digital-to-analog converter (DAC) at room temperature. The readout-in and charge pulses are up-converted to the resonator and qubit frequencies via IQ mixers respectively. The dc-flux bias is applied using a YOKOGAWA GS200 DC source. The control signals are heavily filtered and attenuated before reaching the devices through wiring in the fridge. There is a total of 80 dB attenuation on each readout-in line and 60 dB attenuation on each charge line, all of which are equipped with commercial microwave filters and home-made Eccosorb-filters at the 10~mK stage. For each fast-flux line, we have a total of 30 dB attenuation and a 300 MHz low-pass filter (LPF) followed with an Eccosorb filter at the 10~mK stage. For each dc-flux line, we have a RC filter of 10 KHz cut-off frequency installed at the 4~K stage and a 80 MHz LPF followed by a home-made copper powder filter at the 10~mK stage. Each fast-flux line is combined with dc-flux line via a home-made bias-tee at the 10~mK stage. The readout-output port of the device is connected to an Eccosorb filter, a 5.6-7.0 GHz band-pass filter (BPF) and 3 circulators at the 10 mK stage, protected from the reflected signal and noise coming from the amplifiers. The readout-output signal is amplified by a high electron mobility transistor (HEMT) amplifier at the 4~K stage before further amplification of a room temperature amplifier, and finally sampled by a digitizer.

\section{Processor parameters}

A fluxonium circuit can be modeled by the Hamiltonian
\begin{align}
\hat{H}= 4 E_C \hat{n}^{2}+\frac{E_L}{2}\left(\hat{\varphi}+2\pi \dfrac{\Phi_\mathrm{ext}}{\Phi_0}\right)^{2}-E_J \cos\hat{\varphi} \nonumber, 
\end{align}
where the charge operator $\hat{n}$ and the flux operator $\hat{\varphi}$ are the canonical variables of fluxonium and $E_C$, $E_L$, $E_J$ are its charging, inductive, Josephson energy, respectively. By fitting the Hamiltonian model simultaneously to the measured fluxonium qubit $|0\rangle$ to $|1\rangle$, $\omega_{10}$, and $|0\rangle$ to $|2\rangle$, $\omega_{20}$, transition frequencies versus the external flux $\Phi_\mathrm{ext}$ of each fluxonium, the model parameters are extracted. In addition, the readout resonators are characterized by measuring the resonance frequencies, in the dispersive coupling regime, $\omega_{R,0}$ and $\omega_{R,1}$ for qubit preparing in $|0\rangle$ and $|1\rangle$ states, respectively. The measured device parameters are listed in Table~\ref{table:para}.

\begin{table}
\centering
\begin{tabular}{M{3em} M{3em} M{3em} M{3em} M{3em} M{3em} M{3em}  M{3em}}
\hline \hline 
\multirow{2}{*}{ Qubit } & $E_{C}/h$ & $E_{L}/h$ & $E_{J}/h$ & $\omega_{10}/2\pi$ & $\omega_{21}/2\pi$ & $\omega_{R,0}/2\pi$ & $\chi/2\pi$ \\
&(GHz) & (GHz) & (GHz) & (GHz) & (GHz) & (GHz) & (MHz)\\
\hline
$A$ & $1.398$ & $0.523$ & $2.257$ & $1.09$ & $3.02$ & $6.696$ & $0.63$\\
$B$ & $1.572$ & $0.537$ & $2.086$ & $1.33$ & $3.20$ & $6.753$ & $0.40$\\
\hline \hline
\end{tabular}
\caption{\label{table:para}Device parameters of the two-fluxonium quantum processor.}
\end{table}

\section{Reset \& readout}
\begin{figure}
  \includegraphics{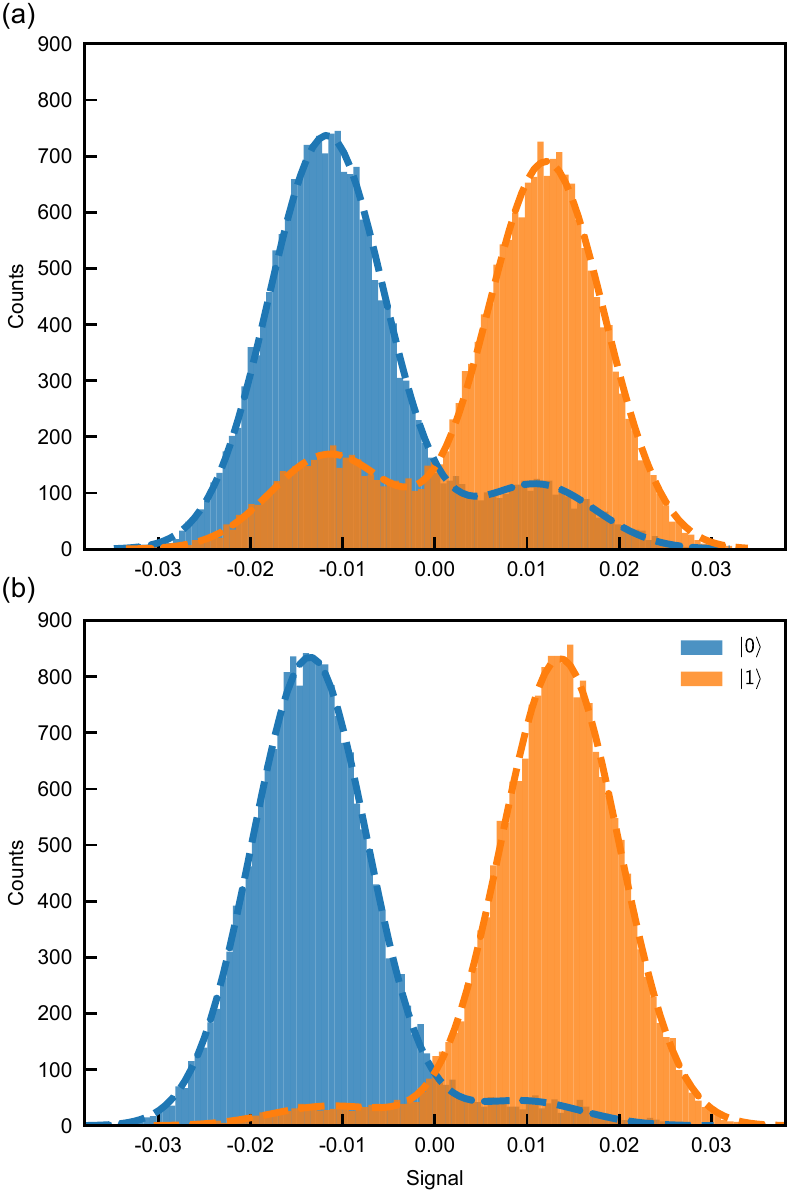}
  \caption{\label{fig:reset_readout}Histograms of the readout signal without (a) and with (b) activate qubit initialization. The dashed line presents the fits to the double-Gaussian model.}
\end{figure}

We use the dispersive readout for qubit state measurement~\cite{wallraff2005approaching}. A near-resonance microwave with 1.5~$\mu$s duration is applied to the readout cavity for homodyne detection. 
Since the qubit has a relatively low transition frequency where $\hbar\omega_{10} \lesssim k_B T$, the qubit is in a thermal mixed state by waiting a long enough time (3-5 times $T_1$). 
In Fig.~\ref{fig:reset_readout}(a), we present the homodyne detection signals of $Q_A$ with the qubit initialized in the thermal mixed state and its population inverted state after a $\pi$-pulse. Because of thermal equilibrium, the qubit always stabilizes at a mixing state. 
The histograms of the in-phase quadrature signal is fitted by a double-Gaussian distribution given as $\sum_i^{0,1}a_i \exp(-(x-x_i)/2\sigma^2)$. The probability $P_i$ of measuring state $|i\rangle$ corresponds to the extracted amplitude $a_i$. For the thermal mixed state, we estimate a residual excitation $P_1=a_1/(a_0+a_1)\approx0.1354$. According to Maxwell-Boltzmann statistics $P_{i}\propto \exp(-E_{i}/k_B T)$, we calculate an effective temperature $T=28$~mK. Here, $E_i$ is the energy of level $i$, $k_B$ is the Boltzmann constant.

To reduce the residual thermal excitation, we utilize the readout resonator as a dissipation channel to reset the qubit. We use a fast-flux pulse to offset qubit away from the flux sweet spot for lifting the selection rule. Simultaneously, a microwave pulse of 20-$\mu$s duration applied to pump the red-sideband transition to transfer the state of the qubit to the resonator. The coupled qubit-resonator system is then reset to the ground states due to the rapid damping of the resonator. In Fig.~\ref{fig:reset_readout}(b), we present the readout distribution of the `ground' and `excited' states after this active qubit reset. The fitting results give an estimation of a residual thermal excitation $P_1=0.0502$. Since other factors like readout imperfections may have increased the measured excitation, the lower bound of reset fidelity is 95\%. With the qubit reset, the readout visibility is increased from 63\% to 88\%. We make corrections to remove the errors due to state preparation and measurement (SPAM) for the population data presented in the main text.

The reset fidelity and readout contrast of $Q_B$ are 96\% and 77\% respectively.

\section{Decoherence}

We measure the coherence times $T_1$, $T_{2,\text{Ramsey}}$, and $T_{2,\text{Echo}}$ of $Q_A$ and $Q_B$ at multiple external flux values near their sweet spots. In the decoherence measurements, the qubit not being measured is placed to its zero-bias point to avoid qubit-qubit interactions. The measurement protocol is similar to that of a tunable transmon qubits as discussed in Ref.~\cite{barends2013coherent} but additional qubit reset is applied to initialize the qubit to the ground state at the beginning of each control-measurement cycle. The qubit is idled at different flux-bias values for decoherence using the high-bandwidth flux line while the qubit control and readout are performed at the flux sweet spot. The coherence times and their flux dependencies are similar between the two qubits. Here, we only present the decoherence data obtained from $Q_A$. 

\begin{figure}
    \includegraphics{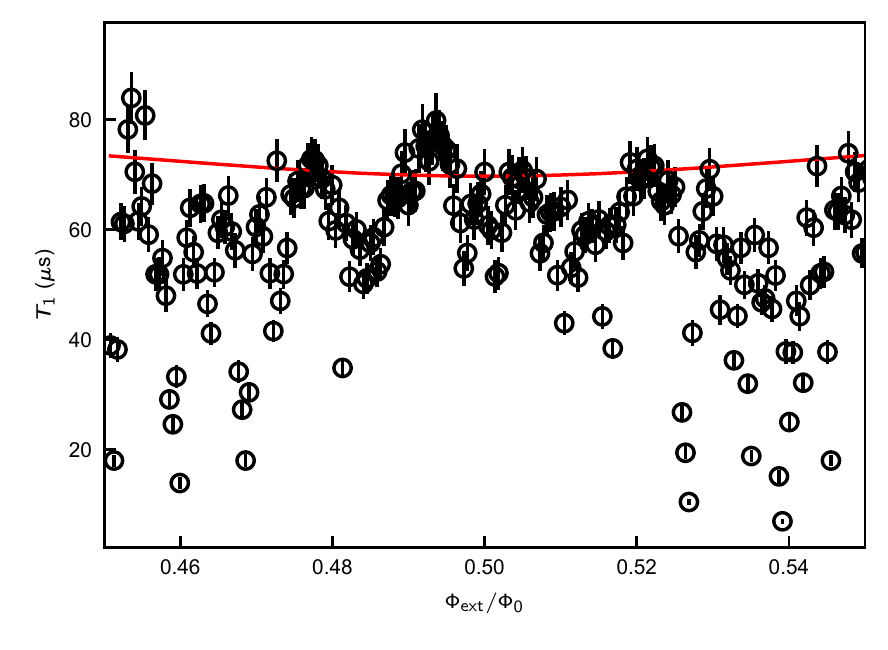}
    \caption{\label{fig:t1_z}Measured relaxation times $T_1$ (black circles) and a prediction from the dielectric loss model (red lines) {\textit{vs.}} the external flux $\Phi_{\mathrm{ext}}$ of $Q_A$.}
\end{figure}

\begin{figure}
    \includegraphics{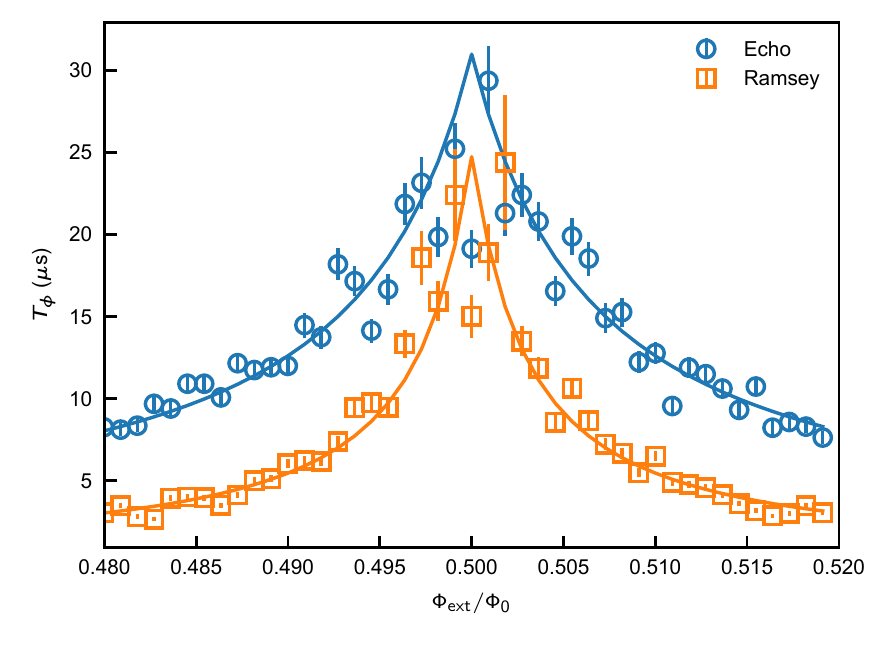}
    \caption{\label{fig:tphi_z}Dephasing times $T_{\phi}$ from the measurements (markers) and those from the prediction from the flux noise model detailed in the main text (lines) {\textit{v.s.}} the $\Phi_{\mathrm{ext}}$ of $Q_A$.}
\end{figure}

\subsection{Energy relaxation}

As shown in Fig.~\ref{fig:t1_z}, $T_1$ can vary between $\sim 20$ to $80$~$\mu$s within a 200~MHz span of qubit frequency around the flux sweet spot. The sharp frequency dependence of $T_1$ is consistent with the behavior of resonant relaxation channels induced by two-level systems (TLS)~\cite{barends2013coherent,klimov2018fluctuations}. Sudden jumps and slow drifts of these resonant relaxation channels versus frequencies, similar to those reported in Ref.~\cite{klimov2018fluctuations, carroll2021dynamics}, are also observed. Apart from the resonant relaxations from the randomly distributed sources at different frequencies, the majority of $T_1$ values lies between 47 $\mu$s and 70 $\mu$s (20-80\% percentile). We attribute this wide-band loss channel to dielectric loss which is also a common loss mechanism for fluxonium~\cite{nguyen2019high}. In Fig~\ref{fig:t1_z}, we show a dielectric loss model (red line) with a loss tangent of $1.7\times 10^{-6}$ and a qubit effective temperature of 30~mK. The dielectric loss model agrees with our data reasonably well and the loss tangent here is comparable with other high-coherence fluxonium demonstrations~\cite{nguyen2019high, somoroff2021millisecond}. 

\subsection{Dephasing}

As shown in Fig.~\ref{fig:tphi_z}, the dephasing times $T_\phi$, measured as characteristic time of the population decay to the $e^{-1}$ fraction of its full scale excluding the relaxation, have a maximum at $\Phi_{\text{ext}} = \Phi_0/2$, known as the flux sweet spot because the qubit is first order insensitive to flux noise at this point. The pure dephasing of the fluxonium qubit away from the sweet spot has a Gaussian rather than exponential decay curve that can be described by the $1/f^{\alpha}$ noise model with $\alpha=1~$\cite{ithier2005decoherence,yoshihara2006decoherence,kou2017fluxonium,nguyen2019high}.
The measured dephasing process can be described by a combination of an energy relaxation rate $\Gamma_1(\Phi_\text{ext})$ (obtained from separate measurements), an external-flux-independent exponential dephasing rate $\Gamma_{\phi1,\beta}$ and an external-flux-dependent Gaussian dephasing rate $\Gamma_{\phi2,\beta}(\Phi_\text{ext})$ as follows:
\begin{equation}
    V_{2,\beta}(t) = A e^{-\frac{\Gamma_1(\Phi_\text{ext})}{2}t  -\Gamma_{\phi_1,\beta}t-\Gamma_{\phi_2,\beta}^2(\Phi_\text{ext})t^2} + B,
\end{equation}
where $\beta$ stands for Ramsey (R) or Echo (E) which refers to values related to Ramsey or spin-echo measurements and $V_{2,\beta}(t)$ is the visibility of the qubit polarization of the $T_2$ measurement. The exponential dephasing $\Gamma_{\phi,1}$ is likely due to white noise sources other than the flux noise. The Gaussian dephasing $\Gamma_{\phi,2}(\Phi_\text{ext})$ is due to the $1/f$ flux noise with noise power spectral density $S(\omega)=A^2/f$. The flux noise amplitude $A$ can be determined from the Gaussian part of the dephasing by
\begin{gather}
    \Gamma_{\phi_2,\mathrm{E}}(\Phiext) = A^{\mathrm{E}}(\ln 2)^{1/2}\left|\partial\omega / \partial\Phiext\right|, \\
    \label{tphi_flux_ramsey}  \Gamma_{\phi_2,\mathrm{R}}(\Phiext) = A^{\mathrm{R}}(\ln (\Gamma_{\phi_2,\mathrm{R}}/2\pi f_{\mathrm{ir}}))^{1/2} \left|\partial\omega / \partial\Phiext\right|,
\end{gather}
for the Ramsey and the spin-echo experiments respectively, where $\partial\omega/\partial\Phiext$ is the flux sensitivity of the qubit transition frequency and $f_{\mathrm{if}}$ is the infrared cutoff of the flux noise. We take $f_{\mathrm{if}}=1$Hz based on our experiment protocol. Note Eq.~\ref{tphi_flux_ramsey} must be solved iteratively. 

The measured dephasing times can be fitted by the above model very well for both the $T_2$ Ramsey and spin-echo measurements. We find the flux noise amplitude $A^{\mathrm{E}}\approx 5.9 ~\mu\Phi_0/\mathrm{Hz}$ at 1 Hz and $A^{\mathrm{R}}\approx ~4.3 \mu\Phi_0/\mathrm{Hz}$ at 1 Hz, measured by the spin-echo and the Ramsey experiment respectively, do not precisely agree with each other. This difference suggests that the $1/f$ noise model is not exact in this device. Introducing a more general form of $1/f^\alpha$ noise with $\alpha \ne 1$ could remedy this discrepancy. 

\subsection{Stability}
\begin{figure*}
    \includegraphics{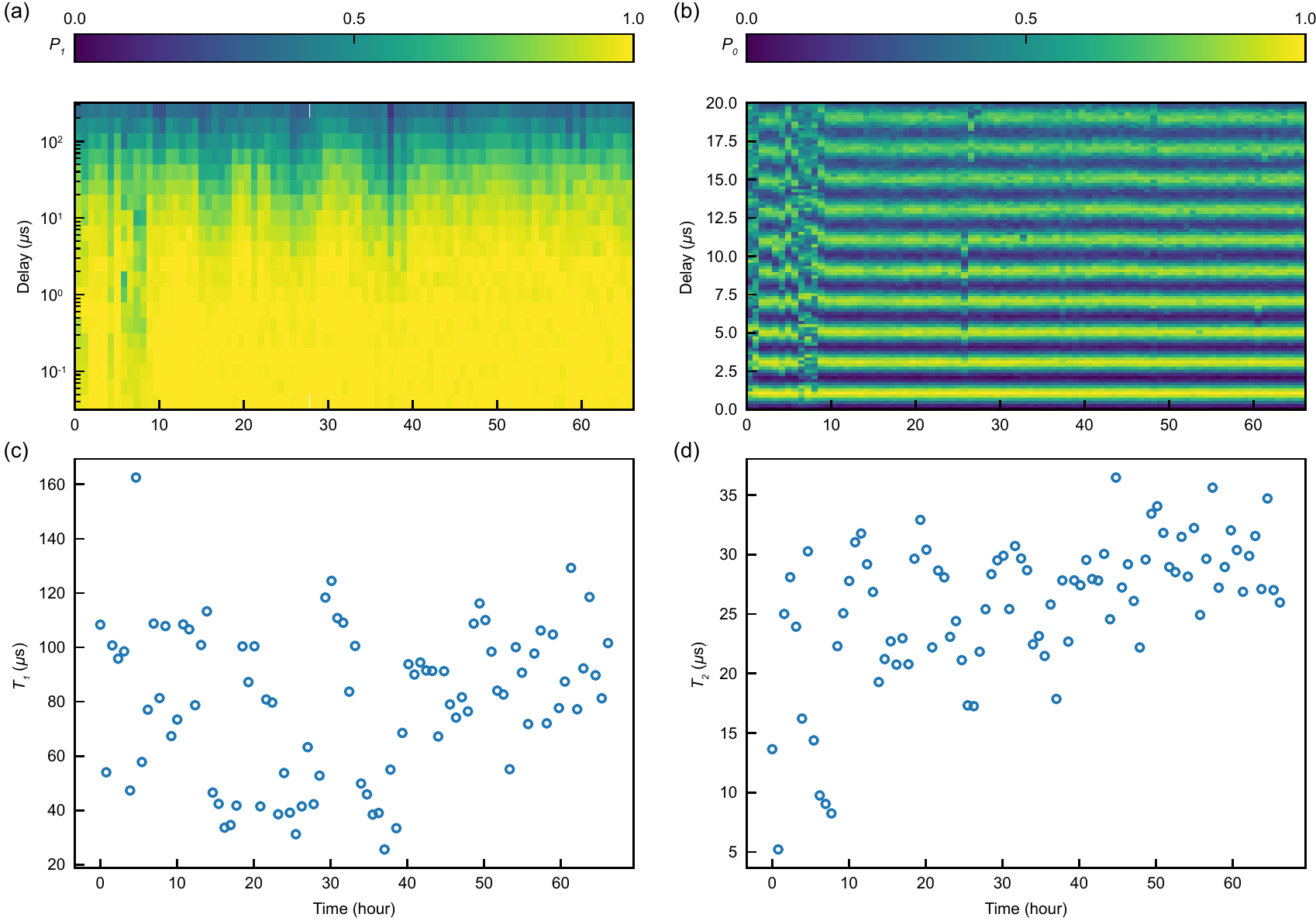}
    \caption{\label{fig:stability}Temporal fluctuations of the energy relaxation time $T_1$ (a,c) and the dephasing time $T_2$ characterized by Ramsey experiments (b,d) of $Q_A$ at $\Phi_\text{ext} = \Phi_0/2$.}
\end{figure*}

We measure the temporal fluctuation of the coherence times $T_1$ and $T_{2,\text{Ramsey}}$ of $Q_A$ for a period spanning a few days, as shown in Fig.~\ref{fig:stability}. The $T_1$ and $T_2$ measurements are executed alternately. During the time of measurement, the coherence times are reasonably long and stable, with $T_1$ fluctuating around 40-120~$\mu$s and $T_{2,\text{Ramsey}}$ around 10-35~$\mu$s. The fluctuations of coherence times are consistent with decoherence due to on-chip fluctuating TLS~\cite{klimov2018fluctuations, burnett2019decoherence}.

\section{Single-qubit gates}

\begin{figure}
  \includegraphics{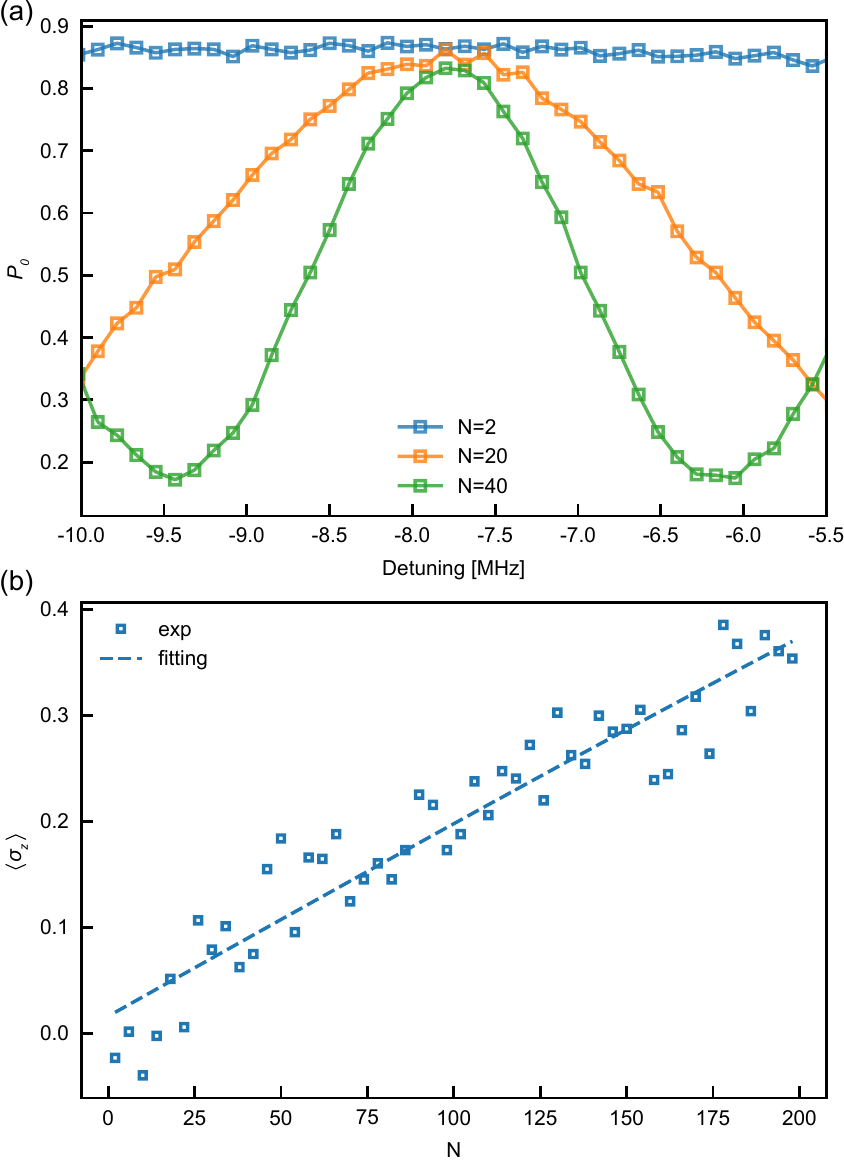}
  \caption{\label{fig:sq_bringup} Calibration of single-qubit gate $X_{\pi}$. (a) Initial state population $P_0$ as a function of detuning for $N=2, 20, 40$ pairs of pseudo-indentities. (b) In the pulse amplitude calibration, $\avg{\sigma_z}$ as a function of the number of $\pi$-rotation.}
\end{figure}

\subsection{Calibration}
We use a cosine envelope for microwave pulses to implement single-qubit gates. To eliminate the ac Stack shift induced by the microwave drive, we introduce a constant detuning from the qubit frequency~\cite{chen2016measuring}. The control pulse is described by
\begin{equation}
\label{pipulse_shape}
\Omega(t) = A \left(1-\cos\left (2\pi \dfrac{t-t_0}{T}\right) \right)e^{-i2\pi \delta f (t-t_0)+\varphi_0},
\end{equation}
where $A$ is the pulse amplitude, $T$ is the gate time, $\delta f$ is the detuning for reducing the phase error, and $\varphi_0$ is used to control the rotation axis. We fix the gate time $T$ and vary the amplitude $A$ to control the rotation angle. 
After the coarse calibrations of $\pi$ and $\pi/2$ rotations around the $X$ and $Y$ axes, we first fine tune the detuning $\delta f$. We use a pseudo-identity operation $X_{\pi}X_{-\pi}$ to amplify the phase error of each gate. Without the phase error, this composite operation constitutes an identity gate which keeps the qubit in its initial state. We sweep $\delta f$ for the different number ($N$) of pseudo-identity operations for a maximum initial state population $P_0$. As presented in Fig.~\ref{fig:sq_bringup}(a), the measured $P_0$ is at its maximum for both $N=2,~ 20~\text{and}~ 40$ when $\delta f$ equals to -7.7~MHz. Since $\delta f$ only weakly depends on the axis of rotation, we use the same detuning parameter for $X(Y)_{\pi}$ gate. The detuning parameter of $X(Y)_{\pi/2}$ gate is calibrated with the same method.

We then fine tune the pulse amplitude $A$ for precise rotation angles. 
Since an incorrect amplitude induces a rotation angle error $\epsilon$, this error is amplified by $n$ times if concatenated $n$ $\pi$-rotations $X_{\pi}$ ($X_{\pi/2}^2$ in the case of calibrating $\pi/2$ rotations) are applied. 
To measure the rotation angle error after amplification, the qubit state vector is prepared alone the $Y$-axis before the $\pi$-rotations. The measurement outcome is a projection of the final state onto the $Z$-axis, described by $\langle \sigma_z \rangle = (-1)^{(n+1)}\sin\left(\left(n+\frac{1}{2}\right) \epsilon\right)$~\cite{sheldon2016characterizing}. In Fig.~\ref{fig:sq_bringup}(b), we present the experiment result of the amplitude calibration of $X_\pi$. The data is fitted well by the above formula and the rotation angle error $\epsilon$ is extracted. The amplitude $A$ should be corrected to $A/(1+\epsilon/\pi)$. We calibrate the pulse amplitude of $\pi/2$ and $\pi$-rotations sequentially.

We next calibrate the misalignment of the rotation axes by adjusting $\varphi_0$. For the $X(Y)$ rotations, the ideal phase is $\varphi_{0,\text{ideal}} = 0\,(\pi/2)$. However, there could be a small misalignment between the ideal and implemented rotation axes. Here, we use the following experiment to measure the axis misalignment, where we use $X_{\pi/2}$ as a reference corresponding to $\varphi_0 = 0$: A first $X_{\pi/2}$ is used to prepare the qubit to a superposition state; The following multiple pairs of operations $(Y_{\pi}X^2_{\pi/2})^N$ rotate the qubit around the $Z$-axis by an angle $-2N(\varphi_{0, ideal} + \delta \varphi_0)$; A final $\pi/2$-rotation with a phase $\pi/2-2N\varphi_{0, ideal}$ is applied to remove the fast population oscillation. In the end, the axis misalignment $\delta\varphi_0$ is extracted by fitting the data with $\langle \sigma_z\rangle=-\sin({2N\delta\varphi_0})$. Similar processes are applied to measure the misalignment of $X_\pi$ and $Y_{\pi/2}$. For $Q_{A}(Q_B)$, the measured $\delta\varphi_0$ of $Y_\pi/2$, $X_\pi$ and $Y_\pi$ are -0.0046(-0.0004)~rad,  -0.0068(0.0455)~rad, and -0.0112(0.0555)~rad, respectively.

\subsection{Randomized Benchmarking}
We further extract the fidelity of calibrated gates with randomized benchmarking (RB) experiments. After randomly applying $m$ Clifford gates to the qubit initialized to its ground state, we measure the average final state $P_0$ of $k$ random sequences as the sequence fidelity $F_{\t{seq}}$. By varying over different choices of $m$, we present the sequence fidelity as a function of $m$ in Fig.~2(a) of the main text. The sequence fidelity of as measured by RB is fitted with a formula $F_{\t{seq}}(m)=A {p_{\t{ref}}}^m + B$, where $A$ and $B$ absorb the SPAM error and $p_{\t{ref}}$ is the decay rate. Since a single-qubit Clifford gate on average consists of 1.875 calibrated primary gates in the gate set $\{I, X_{\pi},Y_{\pi},X_{\pm \pi/2},Y_{\pm \pi/2}\}$, we can estimate an average fidelity of primary gates as $F=1-(1-p_{\t{ref}})(d-1)/d/1.875\approx99.97\%$, where $d=2^{N_q}$ is the dimension of a $N_q$-qubit system. In the interleaved RB, the sequence decay rate $p_g$, the fidelity of the interleaved gate is calculated with $F_g = 1-(1-p_g/p_{\t{ref}})(d-1)/d$~\cite{barends2014superconducting,magesan2012efficient}. The same procedure is also applied to the RB measurement of two-qubit gate fidelities.

\subsection{Single-qubit gate leakage}
\label{subsec:single_qubit_gates_sim}
We perform numerical simulation of the gate dynamics in the fluxonium qubit to estimate the performance of the gates using the QuTiP package~\cite{johansson2012qutip,johansson2013qutip}. The fluxonium Hamiltonian is written in the basis of at least 100 flux eigenstates discretized in $\phi\in [-5\pi,5\pi]$. We include the lowest 5 levels in the time evolution and verified that increasing the number of levels (up to 8) does not affect the results of simulation significantly.

\begin{figure}
    \includegraphics{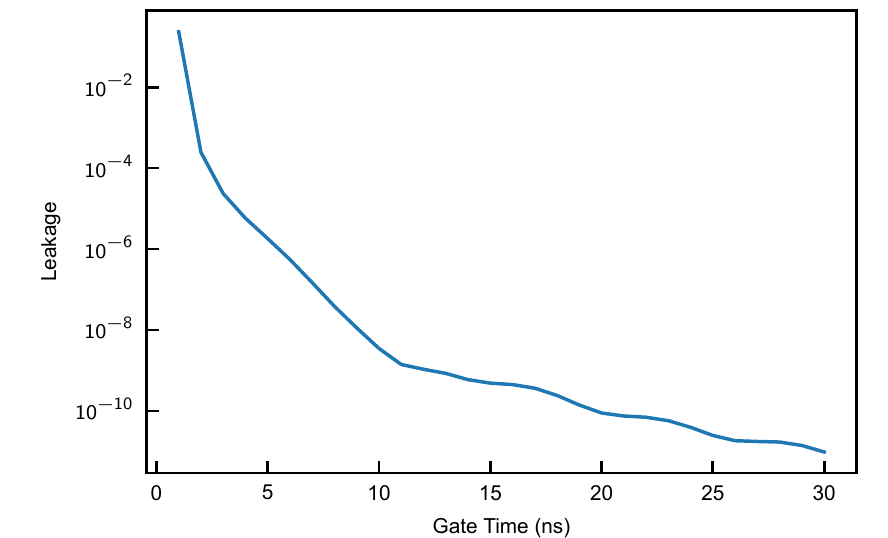}
    \caption{\label{fig:single_qubit_leak}Leakage to the noncomputational levels per $X_\pi$ gate operation from simulations with parameters of $Q_A$.}
\end{figure}

To get the leakage to the non-computational levels of the single-qubit gates, we perform the simulations of the X gate operation by a control pulse defined previously with the driving frequency matching the transition frequency of lowest two levels. We take 2 initial states $|0\rangle$ and $|1\rangle$ and their corresponding final states to construct the evolution matrix $U$ in the basis of $\{|0\rangle,|1\rangle\}$. The leakage of an unitary matrix describing the full-system evolution is defined as the maximum population in the non-computational levels at the end of the operation:
\begin{equation}
    \label{eq:leakage_def}
    r_l = 1 - \min_{|\psi\rangle}\sum_{i=0,1}\left|\langle i|U|\psi\rangle\right|^2,
\end{equation}
where $|\psi\rangle$ is a wavefunction in the computational space.

As shown in Fig~\ref{fig:single_qubit_leak}, due to the large anharmonicity of fluxonium, the leakage decreases with increasing gate time and is below $10^{-5}$ when the pulse is longer than 5~ns, without the need of any further pulse correction. Therefore, decoherence error can be reduced for very short gate time, and the coherence errors can be optimized by adjusting the detuning. This is advantageous to transmon where fine-tuning is required to reduce both the leakage error and total gate error at the same time~\cite{chen2016measuring}.

\subsection{Decoherence error}

\begin{figure}
    \includegraphics{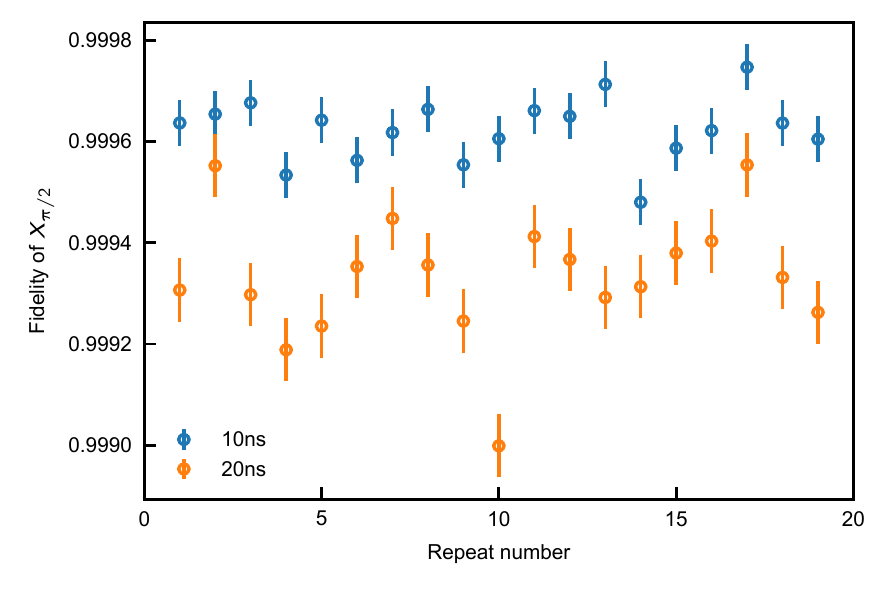}
    \caption{\label{fig:gate_duration}The fidelities of the 10-ns and 20-ns duration $X_{\pi/2}$ gates obtained from repeated RB experiments. The fitting uncertainty is displayed as error bars.}
\end{figure}

We derive the upper bound on fidelity of any single-qubit gate due to decoherence. By definition, average fidelity of the identity gate is given by~\cite{pedersen2007fidelity}
\begin{equation}
    \overline{F}(t, I) = \int d\psi \vert \langle \psi \vert I\cdot U(t) \vert \psi\rangle \vert ^2 = \int d\psi  \langle \psi \vert \psi_f(t)\rangle  \langle \psi_f(t) \vert  \psi\rangle,
\end{equation}
where $U(t)$ is the operator of free evolution, and  $d \psi$ is the Haar measure on state space. $\mathcal{E}_t=  \vert \psi_f(t)\rangle  \langle \psi_f(t) \vert$ is the density matrix of final state $\vert \psi_f(t) \rangle$ after a duration $t$. Under the Bloch-Redfield model of decoherence due to energy decay $T_1$ and dephasing $T_2$ processes, we have 
\begin{equation}
\mathcal{E}_t(\psi) \equiv 
\begin{bmatrix}
1-\vert \beta\vert^2 e^{-t/T_1} & \alpha \beta^* e^{- t/T_2}\\
\alpha^* \beta e^{- t/T_2} & \vert \beta \vert^2 e^{-t/T_1}
\end{bmatrix},
\end{equation}
where $\vert \psi\rangle \equiv \alpha \vert 0\rangle +\beta \vert 1\rangle$. We can now do an explicit calculation
\begin{align}
\overline{F}(\mathcal{E}_t, I) =& \int d\psi \langle \psi \vert \mathcal{E}_t(\psi) \vert \psi\rangle \notag \\ 
=&\int d \psi \vert \alpha \vert^2 + (1- 3\vert \alpha\vert^2 + 2 \vert \alpha \vert^4) e^{-t/T_1}  
\notag \\ &+ 2( \vert \alpha \vert^2 -\vert \alpha \vert^4) e^{-t/T_2} 
\end{align}
We integrate $|\alpha|^2$ and $|\alpha|^4$ over the entire Bloch sphere
\begin{align}
    \int d\psi \vert \alpha \vert ^2 = \frac 1 {4\pi}\int \sin \theta d \theta d \phi \cos^2 \frac \theta 2 = \frac 1 2, \notag \\ 
    \int d\psi \vert \alpha \vert ^4 = \frac 1 {4\pi}\int \sin \theta d \theta d \phi \cos^4 \frac \theta 2 = \frac 1 3. 
\end{align}
Thus,
\begin{equation}
\label{sq_fidelity_upper}
    \overline{F}(\mathcal{E}_t, I) = \frac 1 2 + \frac 1 6 e^{-t/T_1} + \frac 1 3 e^{-t/T_2}.
\end{equation}
Eq.~\ref{sq_fidelity_upper} can be generalized to any single-qubit gate. In the main text, we present the fidelity of $X_{\pi/2}$ and its coherence limit as the function of gate duration. The coherence limit is calculated with $T_1=57~\mu$s and  $T_2=17~\mu$s for $Q_B$. For a specific gate duration, reference RB and interleaved-RB are executed sequentially. We then extract the fidelity of $X_{\pi/2}$ from the pair of measurements. The experiments are repeated for acquiring the average and standard deviation of the gate fidelity. In Fig.~\ref{fig:gate_duration}, we show the repeated measurements of the 10-ns and 20-ns duration gates as examples. The average values of these 19 measurements are shown in the Fig.~2(b) of the main text with the standard deviation displayed as the error bars. The coherence limit of a 10-ns gate is 99.987\% and 99.977\% for $Q_A$ and $Q_B$ respectively. 

\section{Two-qubit gates}

\subsection{Flux crosstalk}
\begin{figure}
  \includegraphics{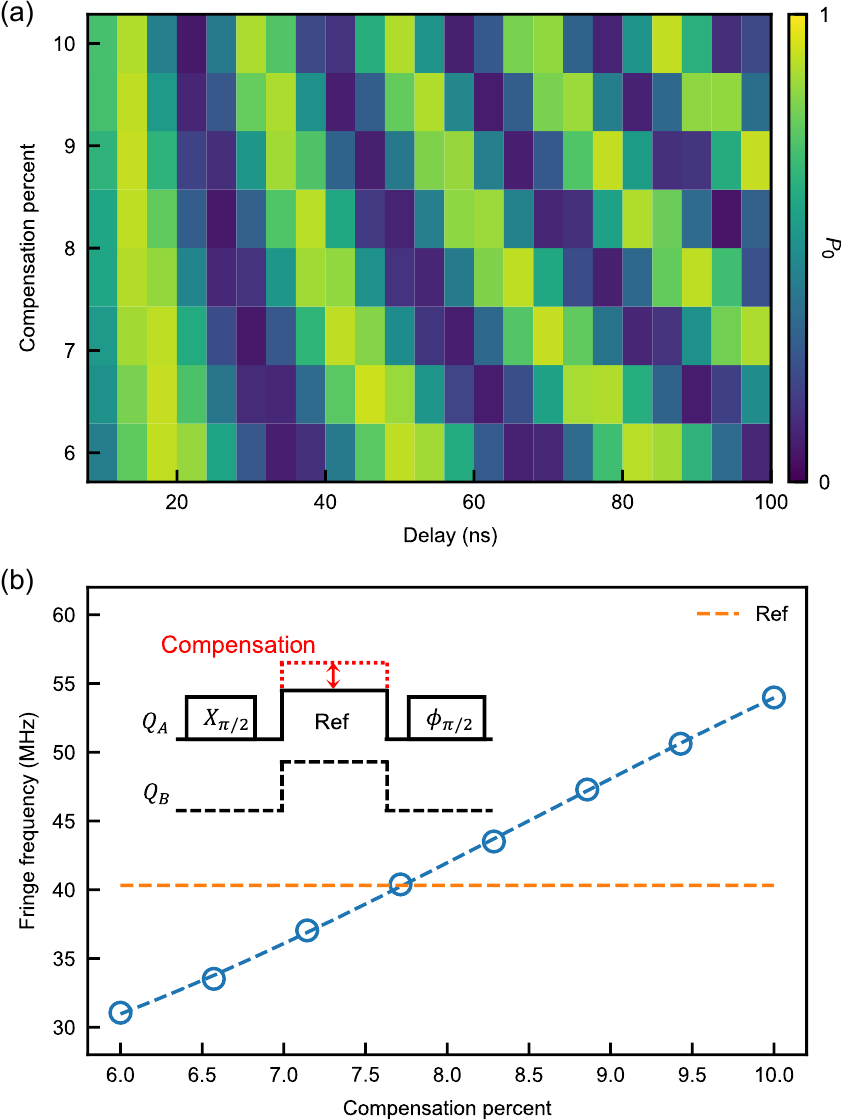}
  \caption{\label{fig:bias_xtalk} Flux crosstalk measurement and compensation. (a) Ramsey oscillations of $Q_A$ as a function of flux compensation (in percentage). (b) The insert shows the control sequence for crosstalk measurement. A Ramsey experiment with a flux pulse on $Q_A$ is used to measure the reference data. After that, a flux pulse is applied to $Q_B$ and generates crosstalk to $Q_A$. Based on the initial flux pulse, an additional compensation pulse is added to mitigating the crosstalk from $Q_B$. The orange dashed line corresponding the oscillation frequency of reference.  The blue circle is execrated from $P_0$ in (a) by the fast-Fourier transformation. The blue dashed line presents the poly-fitting curve. The intersection of two dashed line corresponds to the appropriate compensation.  
  }
\end{figure}

In our device, each fluxonium qubit is controlled via its individual flux line. However, the applied flux not only tunes the frequency of the target qubit, but also acts on the neighbouring qubit. This flux crosstalk should be calibrated before the implementation of the iSWAP gate. We use a standard Ramsey experiment to measure the crosstalk coefficients. As shown in the insert of Fig.~\ref{fig:bias_xtalk}(b), the first $X_{\pi/2}$ is applied to $Q_A$ for preparing a superposition state. After waiting a certain delay, we apply a second half-$\pi$ rotation with a varied phase $\phi$, resulting in a oscillation of the ground state probability $P_0$. The oscillation frequency depends on the qubit frequency. Between the two half-$\pi$ rotations, we put $Q_A$ to a flux sensitive point by a rectangle pulse $z_{ref}$. Without any flux pulse on $Q_B$, we measure the Ramsey oscillation frequency of $Q_A$ as reference. For a flux pulse with a finite amplitude $z_0$ on $Q_B$, the frequency of $Q_A$ is shifted by the crosstalked flux, and can be observed in the Ramsey oscillation. To cancel the flux crosstalk from $Q_B$, we add an additional compensation pulse $z_{com}$ to the $z_{ref}$. In Fig.~\ref{fig:bias_xtalk}(a), we show the Ramsey oscillation as a function of compensation percentage $z_{com}/z_0$. The Ramsey oscillation frequency at varied compensation percentage is exacted and plotted in Fig.~\ref{fig:bias_xtalk}(b). We find an appropriate compensation percent of 7.722\%, when the frequency of $Q_A$ remains unchanged. With similar processes, we can measure a compensation percent 8.375\% for mitigating the flux crosstalk form $Q_A$ to $Q_B$. We end up with the following crosstalk matrix:
\begin{equation}
\begin{bmatrix} V_{\text{in}, A}\\ V_{\text{in}, B} \end{bmatrix} = \begin{bmatrix} 1 & -0.08375 \\ -0.07722 & 1 \end{bmatrix} ^{-1} \cdot \begin{bmatrix} V_{Q, A}\\ V_{Q, B} \end{bmatrix}.
\end{equation}

\subsection{Flux pulse distortion}
\begin{figure}
  \includegraphics{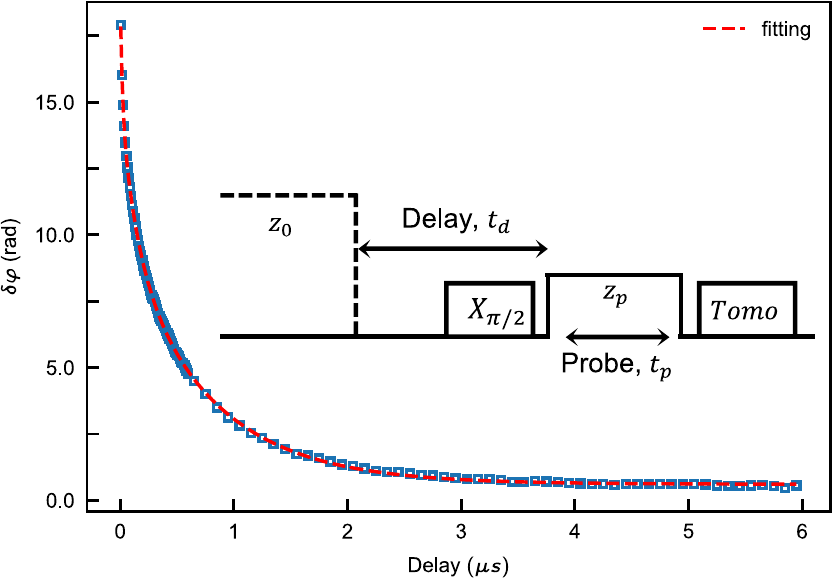}
  \caption{\label{fig:zpulse_distortion}Flux pulse distortion measurement. The phase error $\delta \varphi$ versus time $t_d$ after an initial pulse $z_0$ is shown. The red dashed line represents the fitting curve. The length of initial pulse $z_0$ and the probe pulse $z_p$ are 10~$\mu$s and 50~ns, respectively. The qubit phase $\varphi = \arg (\avg{X}+i\avg{Y})$ is obtained through state tomography.}
\end{figure}

Flux pulse distortion is a common control issue in frequency tunable superconducting qubits, and has been widely studied in many experiments~\cite{rol2020time, sung2021realization}. The distortion is a result of impedance mismatch in the fast-flux lines and can result in an uncontrolled phase error on the qubit. For precise control, the distortion should be measured and compensated accordingly. It has been demonstrated that qubit can be used as an in-situ detector to measure pulse distortion. Usually, the qubit is operated at a flux sensitive point such that flux pulse distortion leads to the qubit frequency deviating from the target value. A Ramsey-like experiment is then used to track the evolution of qubit phase under a distorted pulse and finally the level of pulse distortion can be deconvoluted from the phase. However, qubits suffer from fast dephasing at the flux sensitive points therefore the duration of the qubit evolution is limited by $T_2$ which is in the order of a few microseconds. Here we present an improved version of distortion detection scheme (similar to that in  Ref.~\cite{barends2014superconducting}) which is insensitive to decoherence.

The pulse sequence is shown in the insert of Fig.~\ref{fig:zpulse_distortion}. At the beginning of the sequence, a long and large amplitude reference pulse $z_0$ is applied to the qubit to mimic a step input function at its falling edge. After the falling edge of $z_0$, the system response will continue acting on qubit by inducing its phase variations. A Ramsey-like pulse sequence sequence with a probe pulse $z_p$ on flux sandwiched by two $\pi/2$ rotation pulse is used to probe the phase variation. The flux pulse $z_p$ is for offsetting the qubit to a flux sensitive point. In the absence of any flux pulse, the qubit is idled at its flux insensitive spot. We write down the accumulated phase of qubit as
\begin{equation}
\varphi(t_d) = - \int_{t_d}^{t_d+t_p}\omega_{10}(z(t))\text{d}t,
\end{equation}
where $\omega_{10}(z)$ is the qubit frequency as the function of the flux pulse amplitude $z$, and $t_d$ is the delay between the falling edge and the start of the probe pulse. 
For small distortion, we only consider the first order frequency variation as
\begin{align}
\omega_{10}(z(t)) &= \omega_{10}(z_p) + \left.\dfrac{d\omega_{10}}{dz}\right|_{z=z_p}\cdot z_\text{dist}(t) \notag  \\ 
&= \omega_{10}(z_p) + D(z_p) \cdot z_{\text{dist}}(t)
\end{align}
Here $D(z)$ is the first derivative of qubit frequency with respect to the flux pulse and $z_{\text{dist}}(t)$ is the flux pulse distortion. We also measure the qubit phase without $z_0$ as a reference. The extra phase induced by the distortion is given by
\begin{equation}
\delta \varphi(t_d) = - \int_{t_d}^{t_d+t_p}D(z_p)\cdot z_\text{dist}\text{d}t.
\end{equation}
The experiment data of $\delta \varphi$ is shown in Fig.~\ref{fig:zpulse_distortion}. Here, the distortion can be described by a multi-component exponential function, $z_\text{dist}=z_0 \cdot \sum_i{a_i e^{-t/\tau_i}}$. Therefore the data are fitted with the function 
\begin{equation}
   \delta \varphi(t_d) = z_0 D(z_p)\sum_i \tau_i a_i(e^{-(t_d+ t_p)/\tau_i} - e^{-t_d/\tau_i}). 
\end{equation}
The settling amplitude $a_i$ and settling time $\tau_i$ obtained from fitting are \{-3.52\%,  -1.15\%,  -1.69\%\} and  \{18.5, 143.6, 755.1\}~ns, respectively. With the above parameters, the input pulse pulse is pre-distorted with corresponding impulse response digital filters.

\subsection{$ZZ$ coupling}
\begin{figure}
  \includegraphics{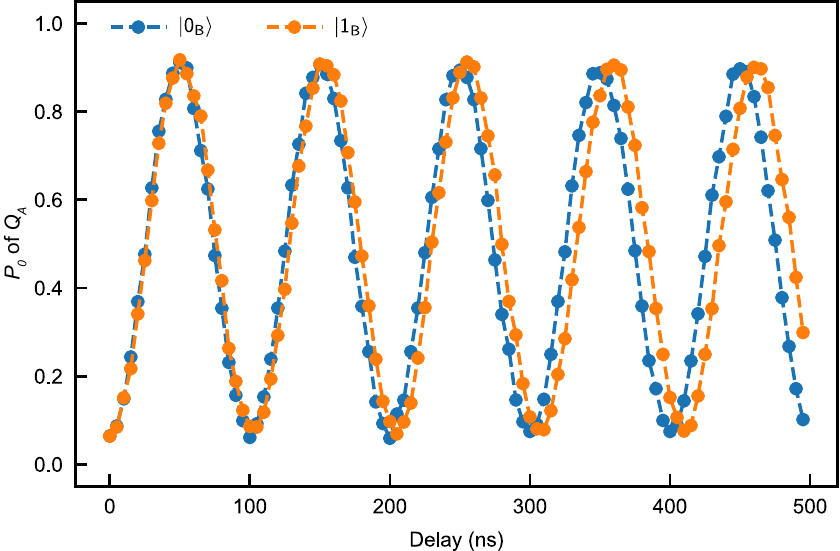}
  \caption{\label{fig:zz_coupling}Measurement of the $ZZ$ coupling strength. The Ramsey oscillation of $Q_A$ under different initial states of $Q_B$.}
\end{figure}

The capacitive coupling between the two qubits leads to an always-on effective $ZZ$ coupling. The $ZZ$ coupling induces a frequency shift of one qubit depending on the state of the other qubit, inducing a controlled-phase error during every operation. Here, we measure the $ZZ$ coupling strength by the standard Ramsey experiment. With $Q_B$ initialed at $|0_B\rangle$ or $|1_B\rangle$, we measure two Ramsey fringe frequencies of $Q_A$. The frequency difference gives an effective $ZZ$ coupling strength of 0.235~MHz (Fig.~\ref{fig:zz_coupling}).

\subsection{Simultaneous single-qubit randomized benchmarking}
\begin{figure}
  \includegraphics{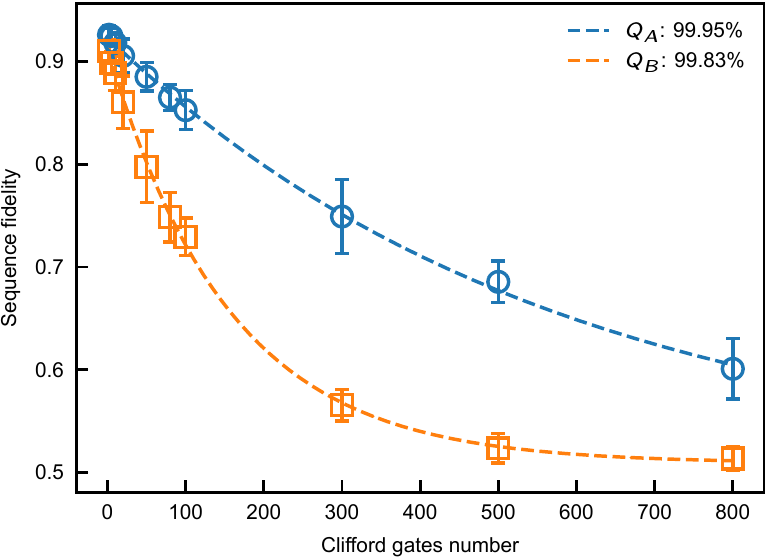}
  \caption{\label{fig:simultaneous_RB}Simultaneous single-qubit RB. The two-qubit sequence fidelity versus the number of single-qubit Cliffords applied simultaneously to both qubits.}
\end{figure}

To quantify the microwave crosstalk, we execute the RB measurement for two qubits simultaneously. The sequence fidelity of single-qubit RB versus Clifford gates number is presented in Fig.~\ref{fig:simultaneous_RB}. For $Q_A$, the fidelity from 99.96\% decreases to 99.95\%. For $Q_B$, the fidelity from 99.84\% decreases to 99.83\%. The single-qubit gate fidelity of $Q_B$ is smaller than that of $Q_A$. The additional error is induced by the on-chip purity loss. The detailed reason is not clear.

\subsection{Calibration of iSWAP gate}
\begin{figure*}
  \includegraphics{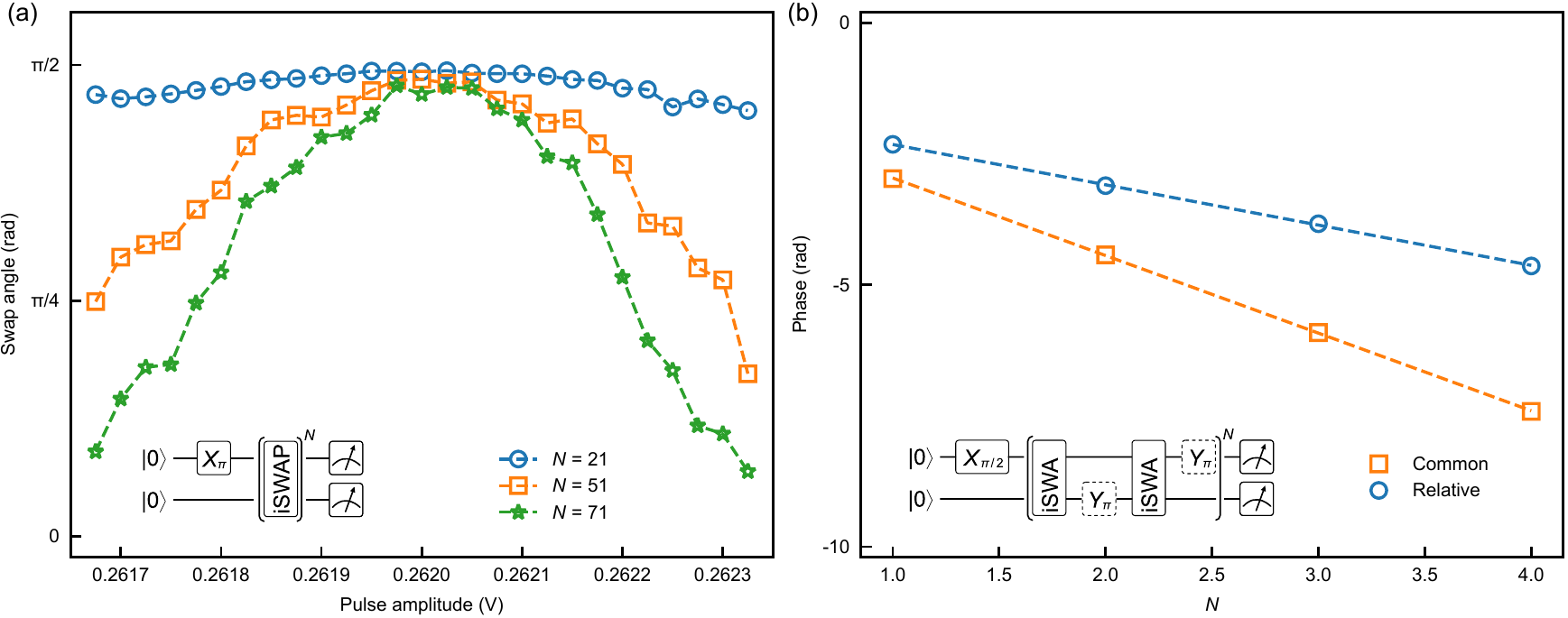}
  \caption{\label{fig:2q_gate} (a) Pulse sequence and measured results for the calibration of the pulse amplitude of the iSWAP gate. (b) Pulse sequence and measured results for the calibration of the relative phase $\chi$ and the common phase $\gamma$ of the iSWAP gate. $Y_\pi$ gates are inserted after each iSWAP gate to amplify and calibrate the relative phase $\chi$.}
 \end{figure*}

As presented in Fig.~3(a) of the main text, we use a flux pulse with error-function shape to bring the $Q_A$ into resonance with $Q_B$. The pulse shape in use is given by
\begin{equation}
    \Phi_{\mathrm{ext}}(t) = \dfrac{\Phi_{\mathrm{amp}}}{2}\left[\mathrm{Erf}(\dfrac{t-4\sigma}{\sqrt{2}\sigma})-\mathrm{Erf}(\dfrac{t-t_g+4\sigma}{\sqrt{2}\sigma}) \right], \label{eq:pulse_shape}
\end{equation}
with $\sigma=0.2$~ns.
At the resonance point, a complete population exchange between $|10\rangle$ and $|01\rangle$ results in an iSWAP gate. Since the energy level of $|11\rangle$ is much lower than that of $|02\rangle$ and $|20\rangle$, the leakage to non-computational levels can be neglected according to simulations. We can simplify the gate operation to a unitary operator in the subspace of $|10\rangle$ and $|01\rangle$ as follows:
\begin{equation}
\label{swap_gate}
U=\left[\begin{matrix}e^{-i\gamma} \cos\theta & -i e^{-i(\gamma-\chi)} \sin\theta &\\ -i e^{-i(\gamma+\chi)} \sin\theta & e^{-i\gamma} \cos\theta \\\end{matrix}\right],
\end{equation}
where $\theta$ is the swap angle, $\gamma$ is the common single-qubit phase, and $\chi$ is the relative phase between $|10\rangle$ and $|01\rangle$. For iSWAP gate, $\theta$, $\gamma$ and $\chi$ are $\pi/2, 0, 0$, respectively. To reduce the error of the swap angle $\theta$, a sequence of concatenated iSWAP gates is used to amplify the swap angle error that can be compensated by adjusting the pulse amplitude. The gate sequence and results are shown in Fig.~\ref{fig:2q_gate}(a). The qubits are prepared in $|10\rangle$ and applied with an odd number of iSWAP gates. The swap angle is calculated from $\theta=\arctan(P_{10}/P_{01})$. We find an optimal pulse amplitude that corresponds to the maximal swap angle. With the optimized pulse amplitude, the swap angle $\theta$ is very close to $\pi/2$, and the diagonal terms in Eq.~\ref{swap_gate} are negligible.

We use a similar method to fine tune the common phase $\gamma$ and the relative phase $\chi$. To calibrate the common phase $\gamma$, the qubits are prepared in a superposition state $(|0\rangle -i|1\rangle)|0\rangle/\sqrt{2}$ followed by an even number of iSWAP gates. The pulse sequence will keep the population in the first qubit but its state evolves into $|0\rangle -ie^{-iN\gamma}|1\rangle$, amplifying the common phase $\gamma$. 
To calibrate the relative phase $\chi$, we insert a $Y_\pi$ gate after every iSWAP gate thus a final state $|0\rangle -ie^{-iN\chi/2}|1\rangle$ is obtained. The gate sequences and results are presented in Fig.~\ref{fig:2q_gate}(b). The final phase of first qubit of the two sequences changes linearly with the number of iSWAP gates, and the fitting slopes are the common phase $\gamma$ and the relative phase $\chi$, respectively. 
Finally, to reduce the effect of pulse distortion, we combine two consecutive flux pulses with opposite amplitudes into a `net zero' pulse~\cite{rol2019fast}. Since the qubits are placed at the sweet spot during idle and other operations, opting for the `net zero' pulse does not affect our calibration procedures. 
After removing these phases with single-qubit phase gates, we can get an iSWAP gate with high fidelity. 

\subsection{Phase correction of iSWAP gate}
For an iSWAP gate, the matrix form of the unitary operator can be written as
\begin{equation}
\label{SWAP_gate}
U_g=\left[\begin{matrix}1 & 0 & 0 & 0\\0 & 0& -i e^{-i(\gamma-\chi)}  & 0\\0 & -i e^{-i(\gamma+\chi)}  & 0 & 0\\0 & 0 & 0 & e^{-i(2\gamma+\phi)}\end{matrix}\right],
\end{equation}
where $\gamma$ is the common phase, $\chi$ is the relative phase, and $\phi$ is the controlled-phase induced by $ZZ$ coupling. In general, the two qubits are parked at different frequencies $\omega_1$ and $\omega_2$ to allow frequency selected microwave control to perform single-qubit gates on each individual qubit. To activate the two-qubit interaction, a fast flux pulse brings the qubit with lower frequency into resonance with the other qubit. An iSWAP gate is implemented in a frame (we call it `exchange frame') in which both qubits are processing at the same frequency. Assuming $\omega_1<\omega_2$, the rotating frame transformation can be written as
\begin{equation}
    R_g=\left[\begin{matrix}1 & 0 & 0 & 0\\0 & e^{- i \omega_2 t} & 0 & 0\\0 & 0 & e^{- i \omega_2 t} & 0\\0 & 0 & 0 & e^{- 2 i \omega_2 t}\end{matrix}\right].
\end{equation}
However, the computational basis in the experimental frame are defined by a different rotating frame transformation
\begin{equation}
    R_o=\left[\begin{matrix}1 & 0 & 0 & 0\\0 & e^{- i \omega_2 t} & 0 & 0\\0 & 0 & e^{- i \omega_1 t} & 0\\0 & 0 & 0 & e^{- i(\omega_1 + \omega_2)t}\end{matrix}\right].
\end{equation}
Hence, the observed gate in the experimental frame which is the single-qubit rotating frame should be described by
\begin{equation}
\begin{aligned}
    U &= R_o^\dagger R_g U_g R_g^\dagger R_o \\ &=\left[\begin{matrix}1 & 0 & 0 & 0\\0 & 0 & -ie^{-i(\gamma-\chi+\Delta t)}  & 0\\0 & -ie^{-i(\gamma+\chi-\Delta t)}  & 0 & 0\\0 & 0 & 0 & e^{-i(2\gamma+\phi)}\end{matrix}\right],
\end{aligned}
\end{equation}
where $\Delta=\omega_1-\omega_2$  is the detuning of the qubits. In the experimental frame, we can observe the dynamical phase induced by the detuning of the qubits. In principle, we can apply a single-qubit phase gate before and after the unitary operator $U$ to remove the unwanted phases. An alternative is to use the virtual $Z$-gate~\cite{mckay2017efficient} to modify the phase of the experimental frame. Here, we explain how we apply frame transformation to perform the phase correction of the iSWAP gate. Supposed we apply the first iSWAP gate at $t=\tau$, if we ignore the controlled-phase $\phi$ , additional phases given by
\begin{equation}
\begin{aligned}
  \varphi_1(\tau) &= \Delta \tau - \chi - \gamma, \\ \notag
  \varphi_2(\tau) &= -\Delta \tau + \chi - \gamma,   
\end{aligned}
\end{equation}
will be added to the quantum state of both qubits. When a subsequent iSWAP gate is applied at $t=\tau^\prime$, not only is a similar additional phase introduced, but the previous phase is swapped
\begin{equation}
\begin{aligned}
    \varphi_1(\tau^\prime) &= \Delta \tau^\prime - \chi - \gamma + \varphi_2(\tau), \\ \notag
    \varphi_2(\tau^\prime) &= -\Delta \tau^\prime + \chi - \gamma + \varphi_1(\tau).  
\end{aligned}
\end{equation}
Hence, to ensure an exact iSWAP implementation, we keep track of the phase difference $\varphi_1$ and $\varphi_2$ between the `exchange frame' and the experimental frame after every iSWAP gate and perform single-qubit phase corrections on all the subsequent signal-qubit rotations with the virtual $Z$-gate.

\subsection{Numerical Simulation of iSWAP gate}

We perform the numerical simulation of the iSWAP gate with each single qubit treated in the same protocol as Sec~\ref{subsec:single_qubit_gates_sim}. Additionally, we construct the $n\times n$ product basis using $n$ lowest single qubit levels of each qubit at the sweet spot for the time evolution.  We pick $n=8$ to account for the hybridization between different levels when the qubit is not at the sweet spot. The flux pulse in use is given by Eq.~\ref{eq:pulse_shape}. As our experiment setup has a low-pass filter on the fast-flux line, we add a Gaussian low-pass filter with a 300~MHz cut-off frequency to mimic the pulse distortion.

To match the experiments, we compute the fidelity of the gate from the simulation~\cite{horodecki1999general,nielsen2002a,nesterov2018microwave} with virtual $Z$-correction:

\begin{align}
    F(\theta_{Aa}&,\theta_{Ab},\theta_{Ba},\theta_{Bb})= \frac{\t{Tr}(U_Z'^{\dagger}U_Z')+|\t{Tr}(U^{\dagger}_{\textrm{iSWAP}}U_Z')|^2}{20} \\
    U_Z  = & \exp(i\theta_{Aa}\sigma_{A,z}\otimes I_2)\exp(i\theta_{Ba}I_1\otimes\sigma_{B,z})U                       \nonumber \\
           & \exp(i\theta_{Ab}\sigma_{A,z}\otimes I_2)\exp(i\theta_{Bb}I_1\otimes\sigma_{B,z}),                                 \\
    F=     & \min_{\theta_{Aa},\theta_{Ab},\theta_{Ba},\theta_{Bb}}F(\theta_{Aa},\theta_{Ab},\theta_{Ba},\theta_{Bb}),
\end{align}
where $U$ is a $4\times 4$ evolution matrix obtained from numerical simulations that started from $\{|00\rangle$,$|10\rangle$,$|01\rangle$,$|11\rangle\}$.

\begin{figure}[htbp]
    \includegraphics{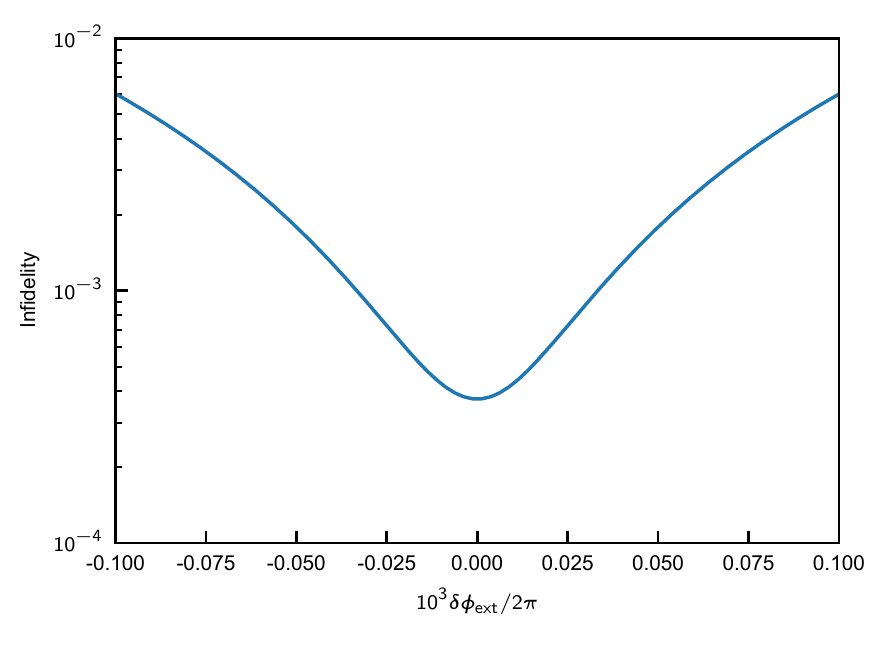}
    \caption{\label{fig:iswap_flux_error}The gate error \textit{v.s.} the flux detuning from the optimal $\Phi_\t{amp}$ of the flux pulse. The virtual $Z$-correction is obtained at the optimal point and then applied to other points to mimic the experiments.}
\end{figure}
As shown in Fig.~\ref{fig:iswap_flux_error}, the iSWAP gate has a minimum error at the optimal point and this error increases very fast when the flux pulse amplitude  moves away from this point. The error at the optimal point comes from the residual $ZZ$ coupling between two qubits and single-qubit excitations due to non-adiabatic flux sweep. We find that the leakage to the noncomputational space is in the order of $10^{-6}$, so it has negligible effect to the gate fidelity.

When the flux pulse amplitude is off the optimal point, the exchange between $|01\rangle$ and $|10\rangle$ is not exact anymore and the angle $\theta$ in Eq.~\ref{swap_gate} deviates from $\pi/2$, which lead to amplitude and phase error. This error is very sensitive to the applied flux and the voltage resolution of the arbitrary waveform generator and actually has a visible effect to the gate fidelity. In our experiments, we estimate the precision of applied flux is approximately $1.5\times10^{-5}\Phi_0$. The coherent error of the iSWAP gate read from Fig.~\ref{fig:iswap_flux_error} is approximately $5\times10^{-4}$.

Besides the coherent error above, we estimate the decoherence error based on the noise model derived in Ref.~\cite{omalley2015qubit}. During the iSWAP gate, $Q_A$ is placed at the resonance point and $Q_B$ stays at the sweet spot. To estimate the error near the resonance point, we assume the $T_{\phi}$ at the sweet spot comes from the white noise and the $T_{\phi}$ near the resonance point comes from the $1/f^{\alpha}$ type flux noise. The rising and falling edges are neglected as they are very short. The total error can be computed as

\begin{align}
    r= & \frac{1}{3}\left[
    \frac{t_g}{T_{1,Q_A,\textrm{r}}} +
    \frac{t_g}{T_{\phi,Q_A,\textrm{sw}}} +
    \frac{t_g}{T_{1,Q_B,\textrm{sw}}} +
    \frac{t_g}{T_{\phi,Q_B,\textrm{sw}}} +
    \frac{t_g^2}{T^2_{\phi,Q_A,\textrm{r}}}
    \right] \nonumber                                                        \\
    =  & 2.8\times10^{-4} + 5.6\times 10^{-4}+ 2.9\times 10^{-4}  \nonumber\ \\
       & + 1.24\times10^{-3} + 7.6\times10^{-4} \nonumber                     \\
    =  & 3.14\times10^{-3}
\end{align}
where $t_g$ is the gate time, $T_{k,Q_i,\textrm{r}}$ is the decoherence time near the resonance point and $T_{k,Q_i,\textrm{sw}}$ is the decoherence time at the sweet spot. $T_{\phi}$ is obtained from $T_1$ and $T_2$ Ramsey measurements. Here we take $t_g=50$~ns, $T_{1,Q_A,\textrm{r}}=60~\mu$s, $T_{\phi,Q_A,\textrm{sw}}=30~\mu$s, $T_{1,Q_B,\textrm{sw}}=57~\mu$s, $T_{\phi,Q_B,\textrm{sw}}=13~\mu$s, $T_{\phi,Q_A,\textrm{r}}=1~\mu$s.

From these calculations we can estimate that the decoherence error accounts for 80\% error in above calculations while the coherent error accounts for the rest 20\%.

\subsection{Decomposition of two-qubit Clifford gates}
Two-qubit Clifford gates can be divided into the following four
classes~\cite{corcoles2013process, barends2014superconducting}:
\begin{enumerate}
    \item $24^2=576$ single-qubit Clifford gates on each individual qubit.
    $$\Qcircuit @C=1em @R=0.7em {
        & \gate{\mathcal{C}_1} & \qw \\
        & \gate{\mathcal{C}_1} & \qw }$$
    \item $24^2\times 3^2=5184$ ``CNOT-like'' gates.
    $$\Qcircuit @C=1em @R=.7em {
        & \gate{\mathcal{C}_1} & \ctrl{1} & \gate{\mathcal{S}_1} & \qw \\
        & \gate{\mathcal{C}_1} & \targ & \gate{\mathcal{S}_1} & \qw }$$
    \item $24^2\times 3^2=5184$ ``$i$SWAP-like'' gates.
    $$\Qcircuit @C=1em @R=.7em {
        & \gate{\mathcal{C}_1} & \push{\boxtimes}\qw & \gate{\mathcal{S}_1} &
        \qw \\
        & \gate{\mathcal{C}_1} & \push{\boxtimes}\qwx\qw & \gate{\mathcal{S}_1}
        & \qw }$$
    \item $24^2=576$ ``SWAP-like'' gates.
    $$\Qcircuit @C=1em @R=.7em {
        & \gate{\mathcal{C}_1} & \qswap \qwx[1] & \qw \\
        & \gate{\mathcal{C}_1} & \qswap & \qw }$$
\end{enumerate}
Here $\mathcal{C}_1=\{C_i, i=1, \cdots, 24\}$ is the single-qubit Clifford
group and $\mathcal{S}_1$ is the group $\{I, S, S^2\}$, where $S =
\exp[-i\pi(X+Y+Z)/3\sqrt{3}]$. In our convention for single qubit Clifford gates, $\mathcal{S}_1 = \{I, Y_{\pi/2} X_{\pi/2}, X_{-\pi/2} Y_{-\pi/2}\}$. In total, there are
$576+5184+5184+576=11520$ two-qubit Clifford gates.

In our experiment, CNOT and SWAP are decomposed to $i$SWAP and single-qubit gates as follows~\cite{schuch2003natural}:
\begin{align*}
    &\Qcircuit @C=1em @R=0em {
        & \ctrl{2} & \qw & & & \push{\boxtimes}\qwx[2]\qw &
        \gate{X_{\pi/2}} &
        \push{\boxtimes}\qwx[2]\qw & \qw \\
        & & & \push{\rightarrow} & & & & &\\
        & \targ & \qw & & & \push{\boxtimes}\qw & \qw &
        \push{\boxtimes}\qw & \qw
    }\\[2em]
    &\Qcircuit @C=1em @R=0em {
        & \qswap \qwx[2] & \qw & & & \push{\boxtimes}\qwx[2]\qw & \qw &
        \push{\boxtimes}\qwx[2]\qw &
        \gate{X_{-\pi/2}} & \push{\boxtimes}\qwx[2]\qw & \qw \\
        & & & \push{\rightarrow} & & & & & & & \\
        & \qswap & \qw & & & \push{\boxtimes}\qw & \gate{X_{-\pi/2}} &
        \push{\boxtimes}\qw &
        \qw & \push{\boxtimes}\qw & \qw
    }
\end{align*}
Symbols in the above circuits are defined as the following.
$$\Qcircuit @C=1em @R=0em {
    & & \ctrl{2} & \qw \\
    \push{\text{CNOT}:} & & & \\
    & & \targ & \qw}\quad
\Qcircuit @C=1em @R=0em {
    & & \push{\boxtimes}\qw\qwx[2] & \qw \\
    \push{i\text{SWAP}:} & & & \\
    & & \push{\boxtimes}\qw & \qw}\quad
\Qcircuit @C=1em @R=0.3em {
    & & \qswap\qwx[2] & \qw \\
    \push{\text{SWAP}:} & & & \\
    & & \qswap & \qw}\quad$$

The error of $C_1$ and $S_1$ can be given by the error of the calibrated single-qubit primary gates, labeled as $r_p$.
\begin{equation}
r_{C_1} = \dfrac{45}{24}r_p,\quad r_{S_1} = \dfrac{5}{3}r_p.
\end{equation}
On average, the error of four classes two-qubit Clifford gates are
\begin{align*}
    r_1 &= \dfrac{45}{24} r_{p, A} + \dfrac{45}{24} r_{p, B}, \notag \\
    r_2 &= \dfrac{109}{24} r_{p, A} + \dfrac{85}{24} r_{p, B} + 2 r_{i\textrm{SWAP}}, \notag \\
    r_3 &= \dfrac{85}{24}r_{p, A} + \dfrac{85}{24} r_{p, B} + r_{i\textrm{SWAP}}, \notag \\
     r_4 &= \dfrac{69}{24}r_{p, A} + \dfrac{69}{24} r_{p, B} + 3 r_{i\textrm{SWAP}}.
\end{align*}
Now, we estimate the average error of an two-qubit Clifford gate $r_{C_2}$ by summing up all the errors as a consistency check,
\begin{align*}
    r_{C_2} &= \dfrac{576}{11520} r_1 + \dfrac{5184}{11520} r_2 + \dfrac{5184}{11520} r_3 + \dfrac{576}{11520} r_4 \notag \\
    &=1.5*r_{i{\textrm{SWAP}}} + 3.875 * r_{p, A} + 3.425 * r_{p, B}.
\end{align*}
According to our RB experiments, the error of the relevant gates are $r_{i{\textrm{SWAP}}}=0.28\%$, $r_{p, A}=0.05\%$ and $r_{p, B}=0.17\%$, respectively. We estimate the error of two qubits Clifford gate $r_{C_2}=1.20\%$, which is very close to the measured value $1.23\%$ from two-qubit RB.

\bibliographystyle{apsrev4-2}
\bibliography{ref}